\documentclass[a4, 11pt]{article}

\usepackage[utf8x]{inputenc}
\usepackage{ucs}
\usepackage{soul,xcolor}
\setstcolor{red}    
\usepackage[round, sort, numbers, authoryear]{natbib}
\usepackage{amsmath}
\usepackage{amsfonts}
\usepackage{amssymb}
\usepackage{latexsym}
\usepackage{graphicx}
\usepackage[position=top]{subfig}
\usepackage{geometry}
\usepackage{lineno}

\usepackage{hyperref}
\hypersetup{
    colorlinks = true,
    urlcolor   = black,
    citecolor  = black,
}

\DeclareTextFontCommand\textsfbi{\usefont{OT1}{phv}{b}{it}}
\DeclareMathAlphabet\mathsfbi            {OT1}{phv}{b}{it}

\newcommand\Tstrut{\rule{0pt}{2.3ex}}         
\newcommand\Bstrut{\rule[-2ex]{0pt}{0pt}}   

\definecolor{myblue}{rgb}{0 0 1}
\definecolor{myred}{rgb}{0.7412 0.2588 0.2588}
\definecolor{mypurple}{rgb}{0.4 0 1}
\usepackage{tikz,siunitx}
\usetikzlibrary{shapes.geometric,shapes.symbols}
\usepackage{gensymb}
\newrobustcmd*{\hexagram}[1]{\tikz{\draw[thick, draw=#1, fill=white] (0cm, 0cm) -- (0.02887cm, -0.05cm) -- (0.08661cm, -0.05cm) -- (0.05723cm, -0.1cm) -- (0.08661cm, -0.15cm) -- (0.02887cm, -0.15cm) -- (0cm, -0.2cm) -- (-0.02887cm, -0.15cm) -- (-0.08661, -0.15) -- (-0.05723, -0.1) -- (-0.08661, -0.05) -- (-0.02887, -0.05) -- cycle;}}
\newrobustcmd*{\emptytriangle}[1]{\tikz{\draw[thick, draw=#1] (0,0) --
(0,-0.18cm) -- (0.1559cm, -0.09cm) -- cycle;}}
\newrobustcmd*{\emptydiamond}[1]{\tikz{\draw[thick, draw=#1] (0,0) --
(0.0684cm,-0.103cm) -- (0cm, -0.205cm) -- (-0.0684cm,-0.103cm) -- cycle;}}

    \makeatletter
\def\@fnsymbol#1{\ensuremath{\ifcase#1\or \dagger\or \ddagger\or
   \mathsection\or \mathparagraph\or \|\or **\or \dagger\dagger
   \or \ddagger\ddagger \else\@ctrerr\fi}}
    \makeatother

\begin{document}
\author
{Khashayar F. Kohan%
  \thanks{Email address for correspondence:
    khashayar.feizbakhshiankohan@mail.mcgill.ca}$\,$ and Susan J. Gaskin\thanks{Email address for correspondence:
    susan.gaskin@mcgill.ca} \\
  Department of Civil Engineering, McGill University,\\ Montr\'{e}al, Qu\'{e}bec, H3A 0C3, Canada}

\date{}

\title{Scalar mixing and entrainment in an axisymmetric jet subjected to external turbulence}

\maketitle

\begin{abstract}
The present study aims to understand the process of turbulent entrainment into a jet, as affected by background turbulence, using scalar statistics. Planar-laser-induced fluorescence was employed to capture the orthogonal cross sections of the jet at a fixed downstream station with varying background turbulence intensities and length scales. The conditional scalar profiles revealed that the thickness of the scalar turbulent/turbulent interface (TTI) is greater than that of the traditional turbulent/non-turbulent interface (TNTI), and the interfacial thickness is an increasing function of the background turbulence intensity. Although nibbling remains the primary entrainment mechanism in the far field, increased occurrence of concentration `holes' within the interfacial layer in the presence of ambient turbulence suggests a more significant role of large-scale engulfment in the turbulent/turbulent entrainment process (although still below 1\% of the total mass flux). Enhanced contribution of the area of detached jet patches (i.e. `islands') to that of the main jet is hypothesized to be evidence of intense detrainment events in the background turbulence. This can potentially result in a reduced net entrainment into the jet, which manifests as less negative values of scalar skewness within the jet core.
\end{abstract}

\section{{Introduction} \label{Introduction}}
Turbulent entrainment refers to the spatio-temporal process by which the ambient fluid, whether irrotational or turbulent, is incorporated into the primary turbulent flow. Conversely, detrainment signifies the transfer of fluid from the primary turbulent flow back into the ambient. The primary turbulent flow is defined as that with the greater turbulence intensity. Entrainment/detrainment are of central importance in numerous phenomena, spanning from bioengineering \citep{Eames&Flor2022} to meteorological events \citep{deRooy2013}. The mechanisms of entrainment, i.e., small-scale viscous/molecular nibbling and large-scale inviscid engulfment, have been studied to a great extent in several turbulent flows in a non-turbulent (quiescent) ambient. For example, in a quiescent background, the nibbling mechanism dominates that of engulfment in jets \citep[e.g.][]{Westerweel2009} and wakes \citep[e.g.][]{Bisset2002}, whereas the contribution of engulfment greatly surpasses that of nibbling in boundary layers \citep{Jahanbakhshi2021}. The properties of the outer layer of turbulent flows in the quiescent ambient, termed the turbulent/non-turbulent interface (TNTI), are also well studied; for a thorough review see \citet{AnnualReview2014}.

Whilst the behavior of the TNTI is well established, relatively few studies have examined the characteristics of the interfacial layer between a turbulent ambient and a turbulent flow, that is, the turbulent/turbulent interface (TTI). The mean effect of background turbulence is to enhance the large- and small-scale undulations of the interface, in what seems to be a universal outcome in wall-bounded \citep{Hancock&Bradshaw1989, You&Zaki2019} and free-shear flows (\citealp{Kohan&Gaskin2022}; \citealp{Chen&Buxton2023}). In other words, the TTI outline is on average rougher than that of the TNTI, where `outline' henceforth denotes the outer boundary of the TNTI and TTI. It is essential to note that albeit thin, the TNTI and TTI have finite thickness, across which the vorticity and scalar adjust between the ambient and the primary turbulent flow.

\begin{figure}
\centering
	\raisebox{1.68in}{(\textit{a})}\includegraphics[width = 0.38\textwidth]{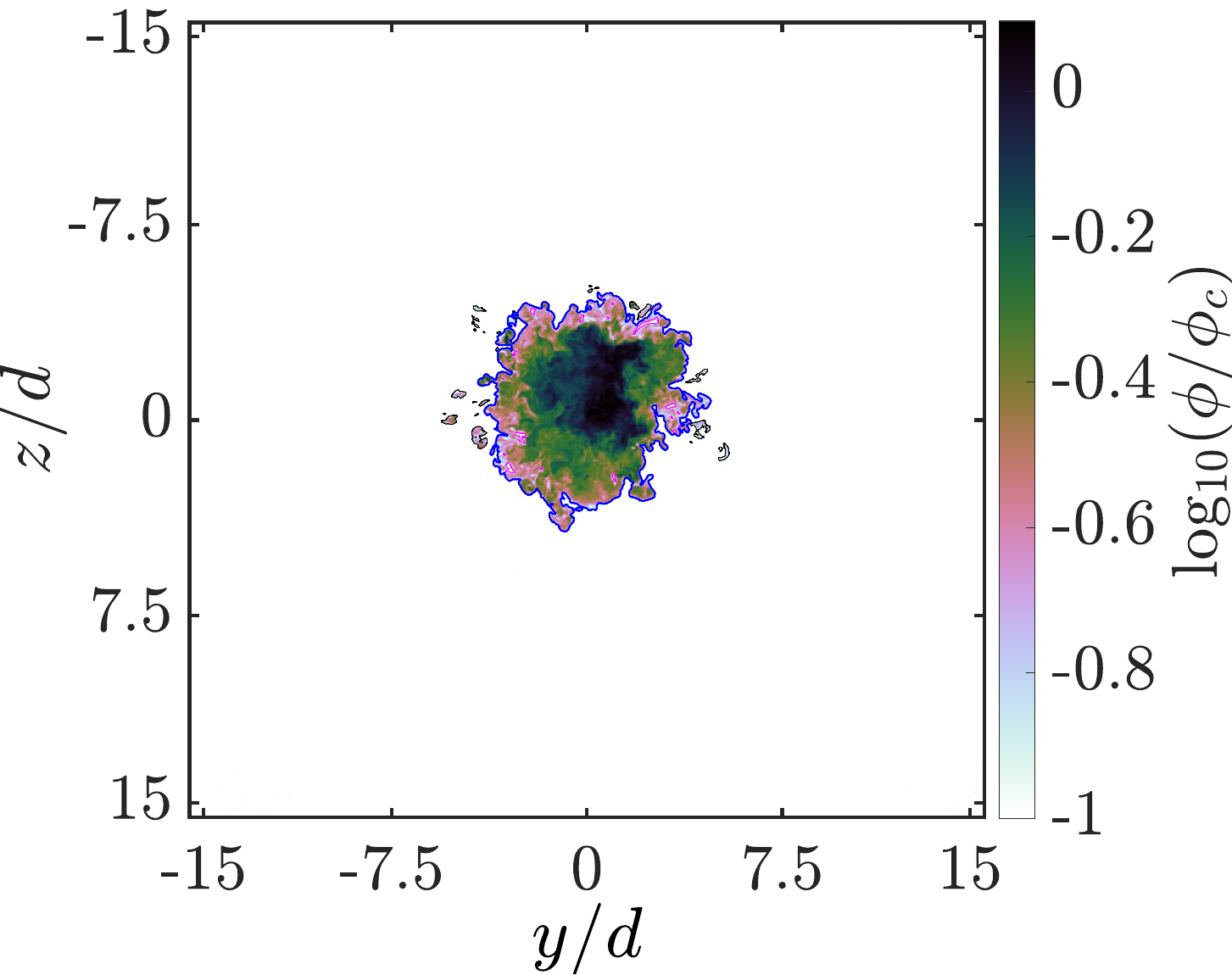}\quad
    \raisebox{1.68in}{(\textit{b})}\includegraphics[width = 0.38\textwidth]{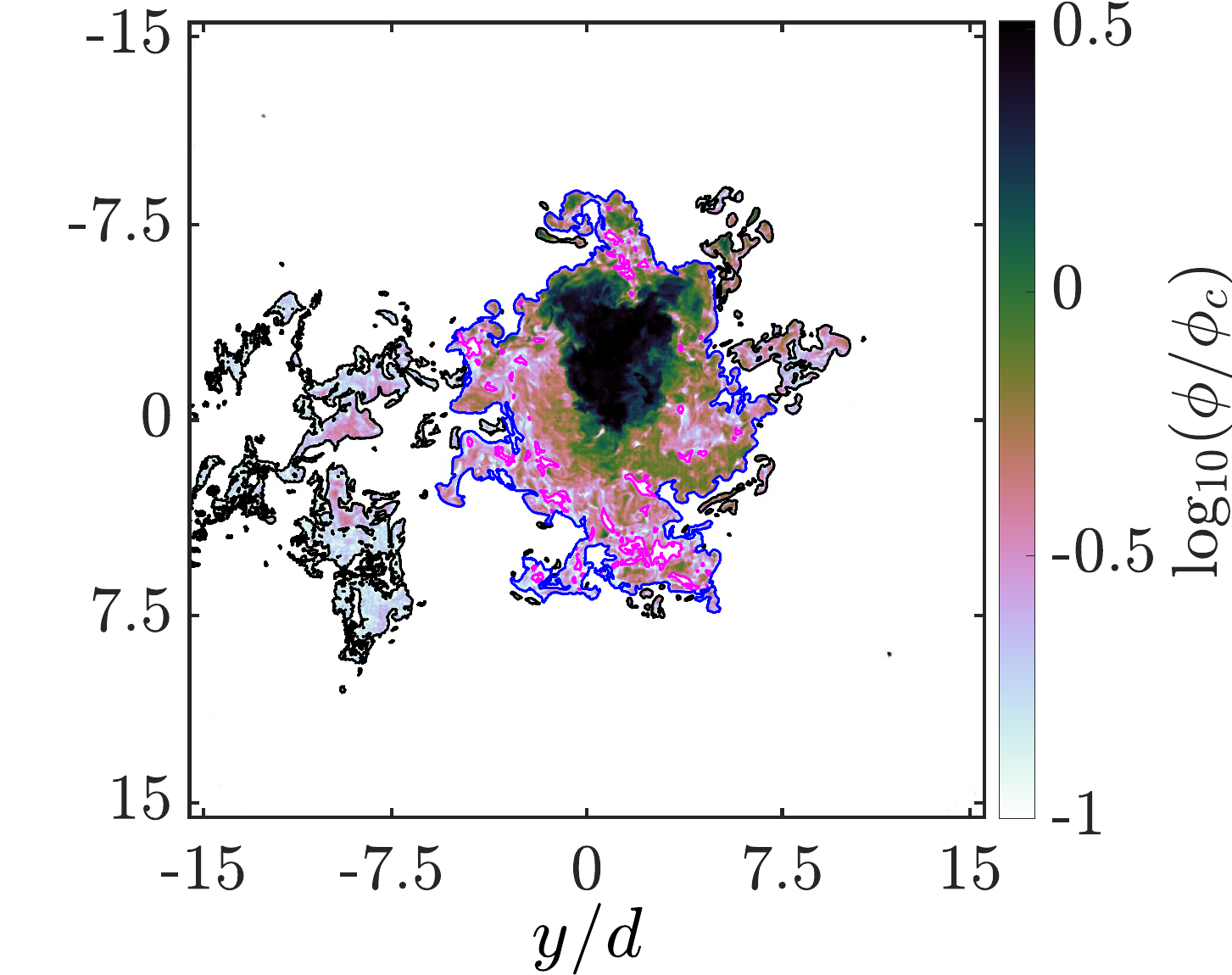}
	\caption{Normalized instantaneous scalar field ($\phi/\phi_c$) of an axisymmetric jet in (\textit{a}) quiescent ambient (case Q) and in (\textit{b}) turbulent ambient (case T3) in logarithmic scaling (Multimedia available online). Here, $\phi_c$ represents the ensemble-averaged centerline concentration. The TNTI and TTI outlines, ambient `holes', and apparently detached `islands' are shown with the blue, magenta, and black lines, respectively. Note the increased presence of islands and holes in the turbulent ambient.}
	\label{Fig.1}
\end{figure}

The presence of external forcing (e.g. turbulence or stratification) may change the basic flow structure, and, therefore, the balance between the entrainment mechanisms. \cite{Westerweel2009} postulated the dominance of engulfment/detrainment over nibbling in free-shear flows exposed to strong external forcing. Intermittent detrainment events have been observed in experiments involving filling-box plumes attached to a vertical wall and subjected to sufficiently strong ambient stratification. This phenomenon, termed `plume peeling', manifests as intrusions of passive scalar patches originating from the plume into the environment (e.g. \citealp{Gladstone&Woods2014}; \citealp{Bonnebaigt2016}). An increased occurrence of detrainment events was also qualitatively demonstrated for boundary layers in external turbulence relative to a non-turbulent ambient \citep{You&Zaki2019}, consistent with our visualizations in figure \ref{Fig.1} (Multimedia available online). Recent experimental measurements of \cite{Kankanwadi&Buxton2023} revealed that grid-generated background turbulence enhances the net rate of entrainment into the near field of cylindrical wakes, where large-scale coherent structures (i.e. von K\'arm\'an vortex street), and, thus, engulfment, prevail. This result, which is in direct contrast to the decrease in entrainment due to nibbling of the far field, implies that free-stream turbulence enhances large-scale engulfment in wakes.

The effect of background turbulence on the entrainment process in the far field of a jet is three-fold in the sense that it acts (i) to increase the surface area across which entrainment occurs \citep[e.g.][also see figure \ref{Fig.1}]{Kohan&Gaskin2022}, (ii) to break up the coherent structures of the jet, thereby reducing the mean axial velocity and the induced entrainment wind \citep{Hunt1994, Khorsandi2013}, and (iii) to promote large detrainment events as compared to the non-turbulent background. It is generally understood that the second and third effects dominate the former, and, thus, the rate of entrainment is suppressed in a turbulent ambient (\citealp{Hunt1994}; \citealp{Gaskin2004}; \citealp{Khorsandi2013}; \citealp{Lai2019}; \citealp{Sahebjam2022}). Eventually, the turbulent ambient breaks up the jet at a critical background turbulence intensity. For example, in a zero-mean-flow approximately homogeneous background turbulence generated by a random jet array (RJA), the onset of break-up for an axisymmetric jet occurs once $\xi = u_{\tau}/u_{jet, q}^{rms} > 0.5$ \citep{Sahebjam2022}, where $\xi$, $u_{\tau}$, and $u_{jet, q}^{rms}$ denote the relative turbulence intensity between the ambient and the jet, the characteristic velocity of the ambient turbulence, and streamwise ($x$) root-mean-square (r.m.s.) velocity at the jet centerline in the quiescent ambient, respectively. The jet-driven entrainment halts beyond the break-up point, and the TTI outline diffuses like a passive interface in the turbulent ambient \citep{Hunt2006}. In light of these observations, however, it becomes evident that understanding of the impact of external turbulence on the relative importance of different entrainment mechanisms into the jet is still lacking.

Our intention is to provide new insights on the interplay between nibbling, engulfment, and detrainment of jets in external turbulence. Hence, in the present study, we investigate the interfacial and entrainment processes in the far field of an axisymmetric jet subjected to RJA-generated turbulence, using scalar statistics. The methodology is briefly presented in \S\ref{Methodology}, while the main findings are demonstrated in \S\ref{Results}. The conclusions are drawn in \S\ref{Conclusions}.


\section{Methodology \label{Methodology}}
 The planar laser-induced fluorescence (PLIF) measurements were carried out in a $1.5 \times 6 \times 1\, \textrm{m}^3$ open-top glass tank with walls of tempered glass to provide optical access. A 12-bit, 4 MP CMOS camera (pco.dimax), fitted with a low-pass filter, was used at a sampling frequency of 50 Hz to capture the scalar field in orthogonal cross sections of a round jet at a downstream station of $x/d = 25$ for a total of 1500 to 3000 instantaneous fields for the investigated cases \citep[]{Kohan&Gaskin2022}. The jet exited parallel to the plane of the RJA, having a Reynolds number of $Re_J = 10600$ based on its diameter ($d = 8.51$ mm) and exit velocity. The field of view (FOV), spanning a region of $260 \times 260\, \textrm{mm}^2$, was illuminated by a 1.5 mm thick laser sheet, which was formed by an 8-sided polygonal rotating mirror. In order for the scalar to faithfully track the jet fluid (water), a high-Schmidt number ($Sc \gg 1$) passive scalar was used, namely, Rhodamine 6G with $Sc \approx 2500$. The current PLIF resolution (assessed as $\approx 1.7$ centerline Kolmogorov microscale in the quiescent ambient, $\eta_q$) is sufficient to capture the conditional scalar statistics relative to the location of the outline \citep{Kohan&Gaskin2020}, which is required to estimate the thickness of the interfacial layers. The jet-centerline values of the Kolmogorov and Taylor ($\lambda_q$) microscales in the quiescent ambient are calculated from the empirical relations of \cite{Friehe1971},

\refstepcounter{equation}
$$
  \eta_q/d = (48 Re_J^{3})^{-1/4}(x/d), \quad
  \lambda_q/d = 0.88 Re_J^{-1/2}(x/d).
  \eqno{(\theequation{\mathit{a},\mathit{b}})}\label{Eq.1}
$$

\begin{figure}
	\raisebox{1.73in}{(\textit{a})}\includegraphics[width = 0.323\textwidth]{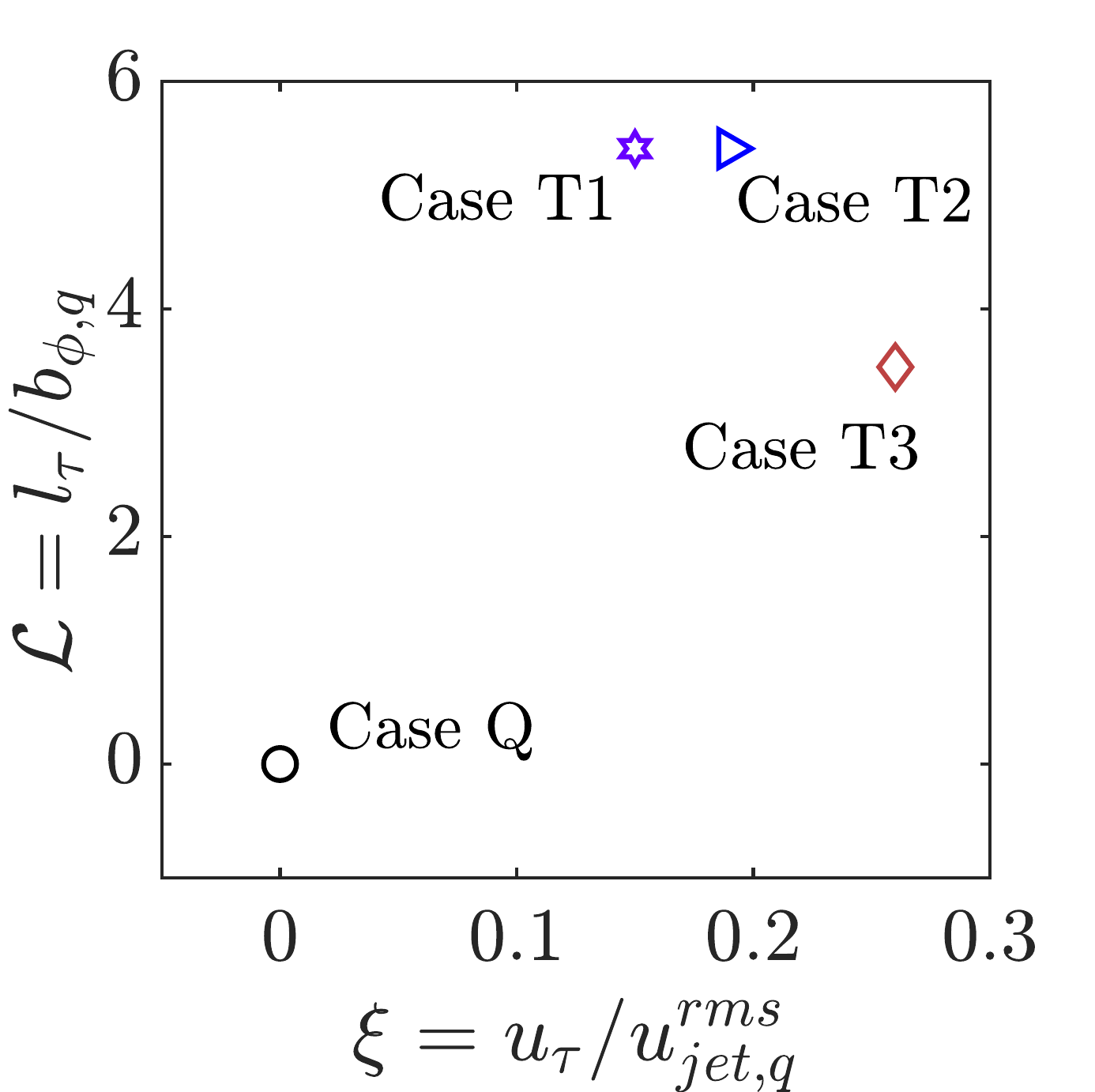}\hspace*{0.2cm}
    \raisebox{1.73in}{(\textit{b})}\includegraphics[width = 0.65\textwidth]{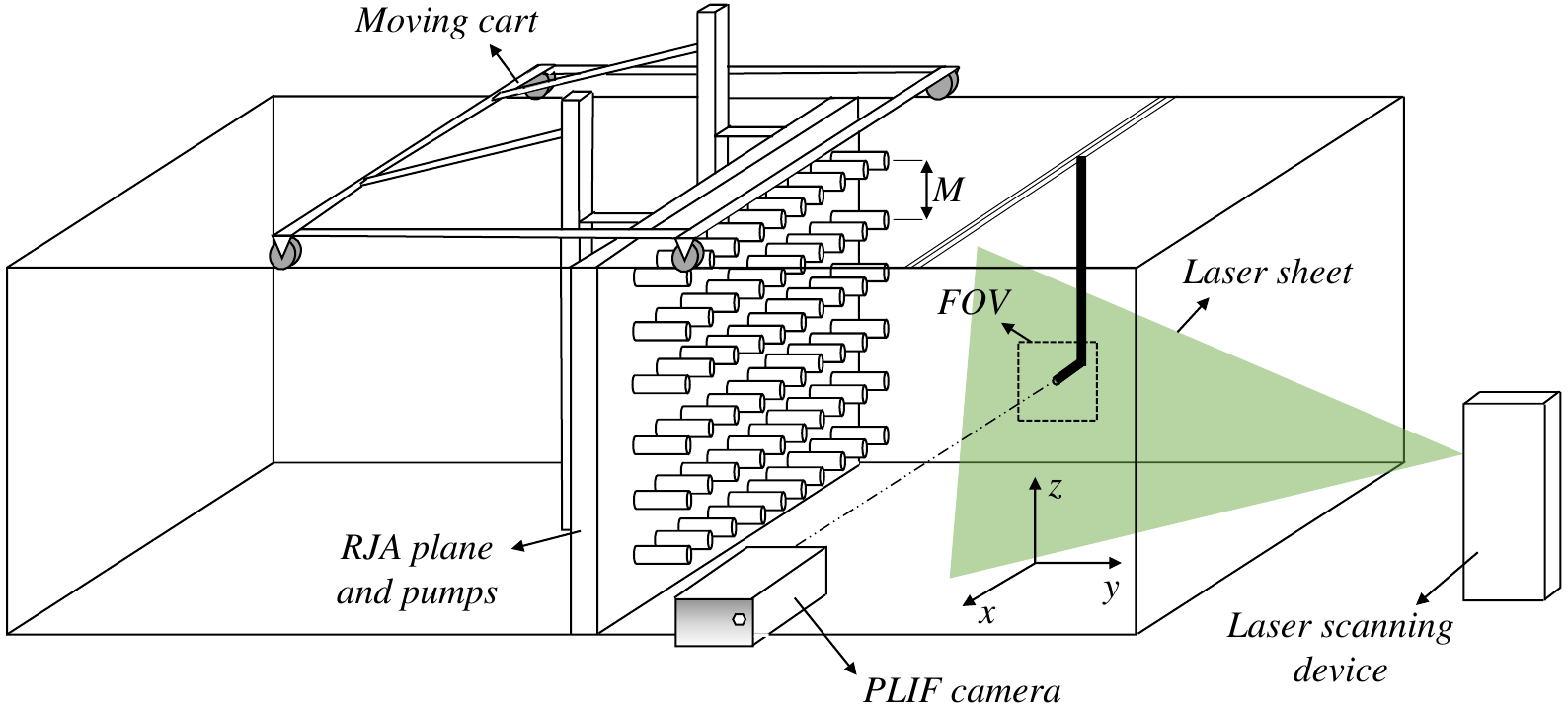}
	\caption{(\textit{a}) Experimental envelope presenting the relative turbulence intensity and length scale of the conducted cases. (\textit{b}) An illustration of the experimental set-up.}
	\label{Fig.2}
\end{figure}

 A systematic study was conducted to elicit the effect of increasing background turbulence intensity on the properties of the interfacial layer and the entrainment process. We note that the length scale of the ambient turbulence, $l_{\tau}$, is relatively unimportant in characterizing the behavior of the interfacial layer in the far field for the range of $l_{\tau}$ investigated herein \citep[][]{Kohan&Gaskin2022}. Nonetheless, figure \ref{Fig.2}(\textit{a}) presents the relative turbulence intensity, $\xi = u_{\tau}/u^{rms}_{jet,q}$, and length scale, $\mathcal{L} = l_{\tau}/b_{\phi,q}$, between the ambient and the jet for the studied cases, where $b_{\phi,q}$ represents the concentration half-width of the reference case, that is, the jet in a quiescent background (case Q). The other runs in this experimental campaign are named such that the prefix (case $\underline{\textnormal{T}}\#$) denotes the jet in the turbulent ambient, while the hierarchical order of the suffix (case $\textnormal{T}\underline{\#}$) reflects the increasing background turbulence intensity. Further details on the calculation of $\xi$ and $\mathcal{L}$ can be found in \cite{Kohan&Gaskin2022}.

The characteristics (i.e. $u_{\tau}$ and $l_{\tau}$) of the zero-mean-flow external turbulence were controlled by moving the RJA sheet relative to the jet exit (along the $y$-axis); the closer the RJA to the jet, the more intense the background turbulence. The RJA comprised 60 bilge pumps with center-to-center distance of $M = 15$ cm (figure \ref{Fig.2}\textit{b}). The optimized `random' algorithm \citep[][]{Perez-Alvarado2016} driving the RJA, generated an approximately homogeneous turbulence across the average width of the jet for the cases considered here. In this `random' algorithm, the pump on/off times followed two normal distributions with parameters ($\mu_{on},\sigma_{on}$) = ($12,\, 4$) s and ($\mu_{off},\sigma_{off}$) = ($108,\, 36$) s, resulting in 10\% of the pumps being on at any instant (on average). The unavoidable decay of the RJA-generated turbulence does not systematically influence the behavior of the jet across its width, and, thus, averaging the scalar statistics over the TTI outline length is considered appropriate \citep{Kohan&Gaskin2022}.

\section{Results \label{Results}}
\subsection{Scaling of the scalar TNTI and TTI \label{TNTI and TTI thickness}}

The identification of the TNTI and TTI outlines is performed by placing a threshold, $\phi_t$, on the instantaneous scalar concentration fields, noting that $0.11 \leqslant \phi_t/\phi_c \leqslant 0.14$ for the cases considered herein. Thereafter, the conditional profiles (denoted by $\langle \sim \rangle_I$) are assessed as ensemble-averaged flow variables (i.e. mean and r.m.s. concentration) along the local interface-normal coordinate, represented by $x_n$. Note that $x_n > 0$ and $x_n < 0$ point into the jet and into the ambient, respectively, while $x_n = 0$ lies on the outline.

The existence of the scalar TTI was first verified in Kohan \& Gaskin (2022) with the evidence reproduced here in the form of figure \ref{Fig.3}(\textit{a}). This figure reports the behavior of conditional mean scalar relative to the position of the outline, $\langle \phi \rangle_I$, exhibiting sharp jumps across the TTI layers akin to the classical TNTI. It is also evident that the value of the conditional jump monotonically increases with $\xi$. This suggests the possibility of enhanced transport of concentration from the jet core towards the edges in background turbulence, causing greater levels of passive scalar to exist within the interfacial layer.

\begin{figure}
\centering
\raisebox{1.75in}{(\textit{a})}\includegraphics[width = 0.328\textwidth]{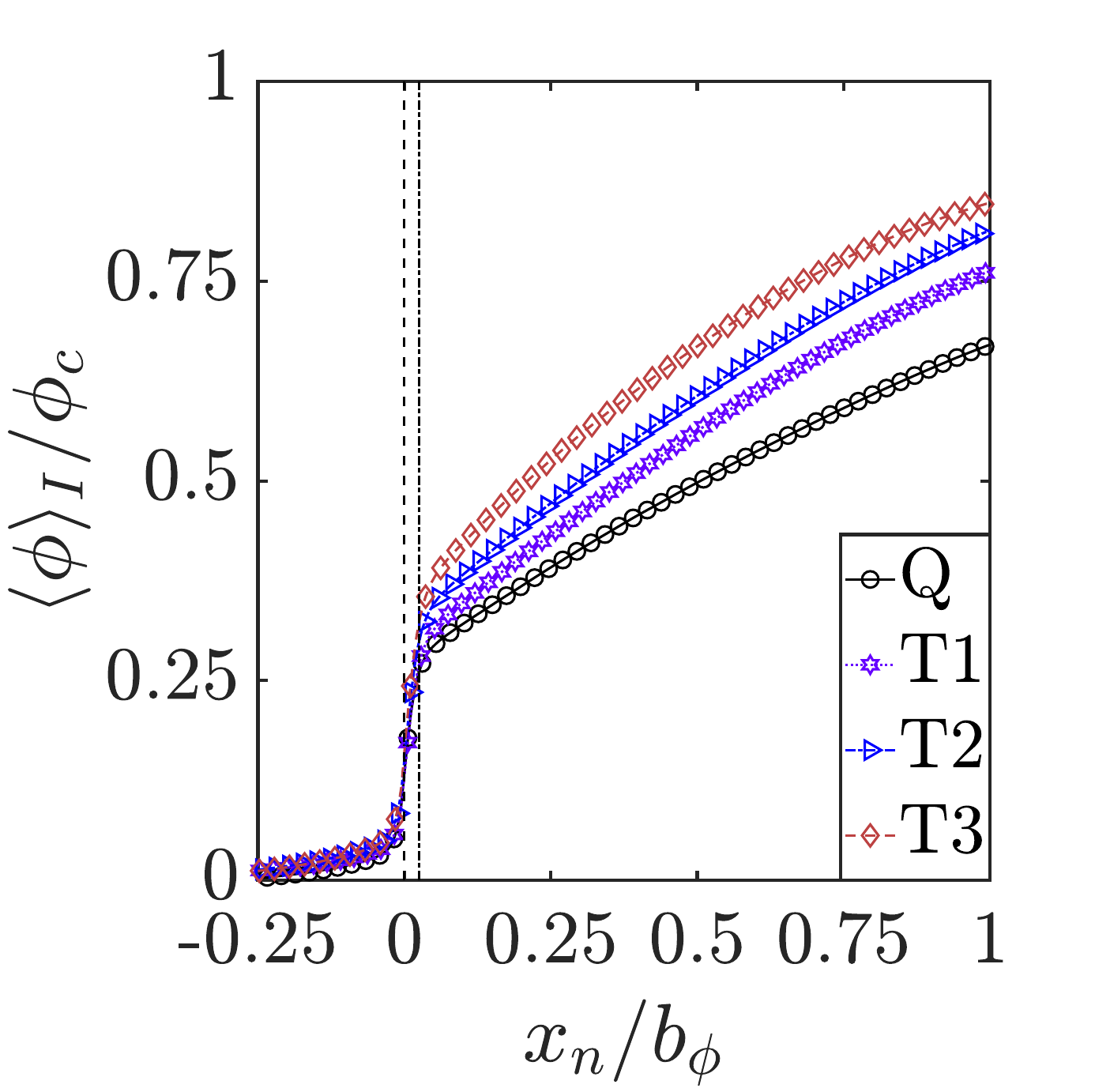}\quad
\raisebox{1.75in}{(\textit{b})}\includegraphics[width = 0.328\textwidth]{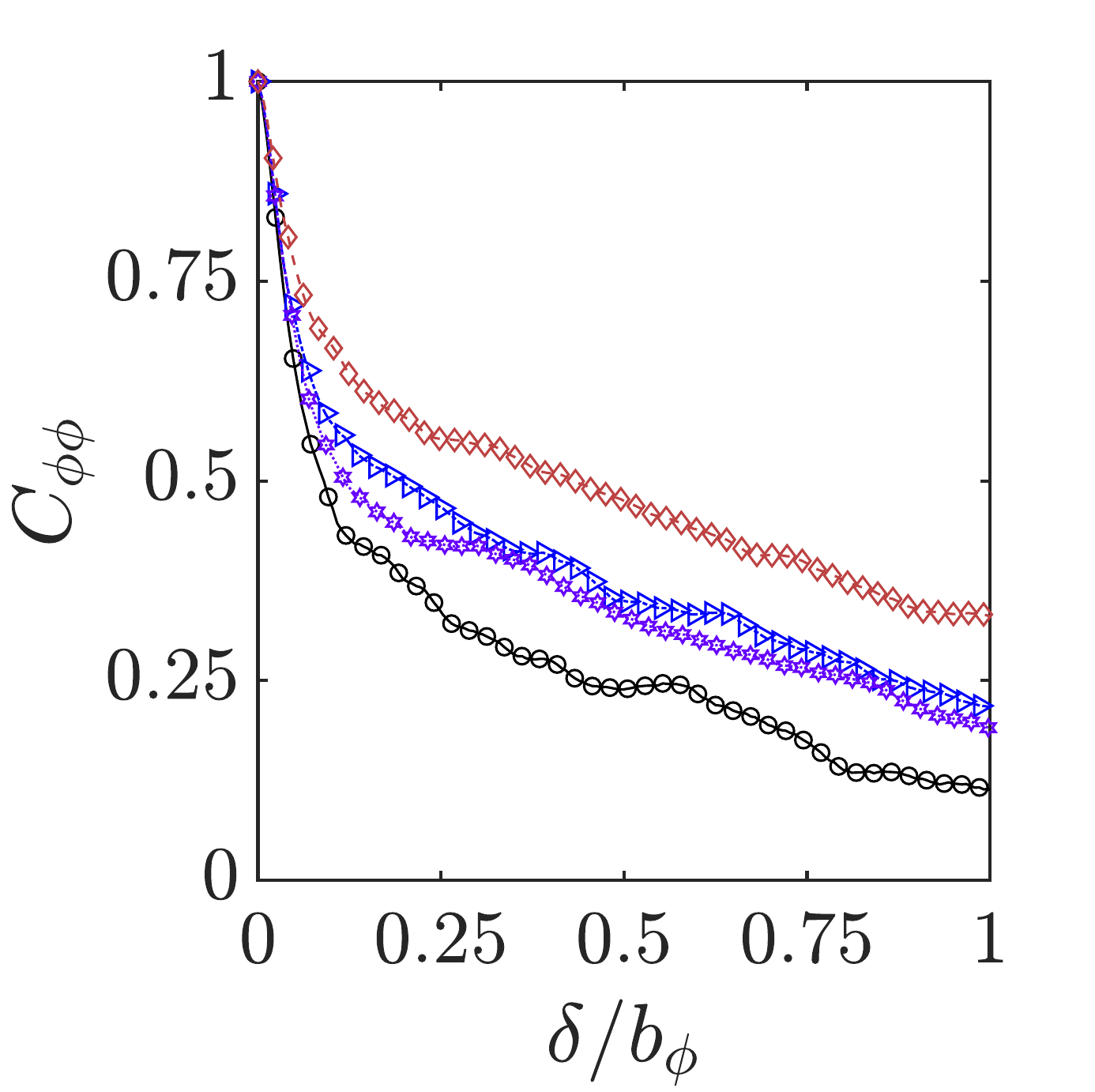}\\
\raisebox{2.76in}{(\textit{c})} \hspace*{-0.43cm}\raisebox{1.32in}{(\textit{d})}\includegraphics[width = 0.66\textwidth]{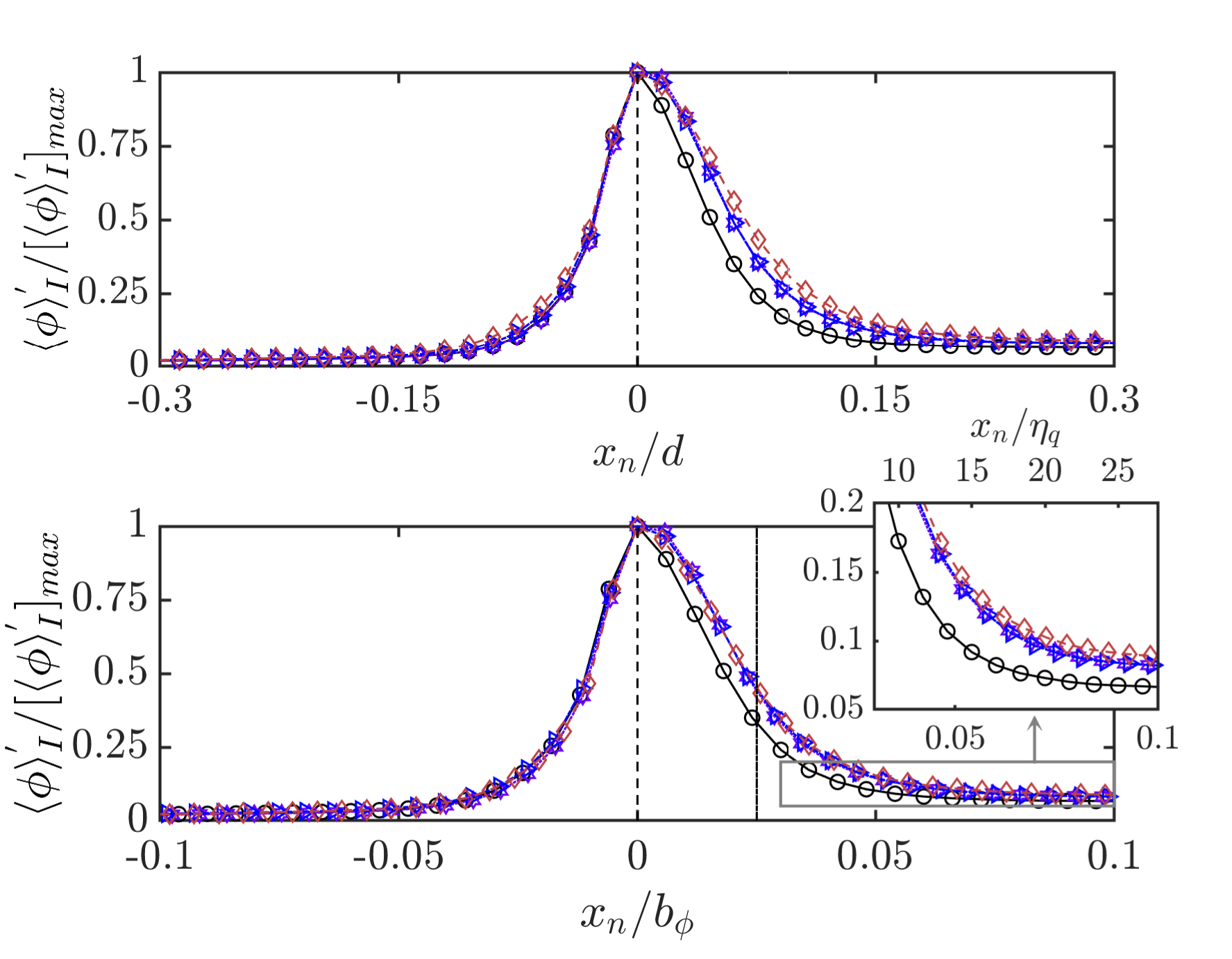}
\raisebox{1.77in}{(\textit{e})}\includegraphics[width = 0.328\textwidth]{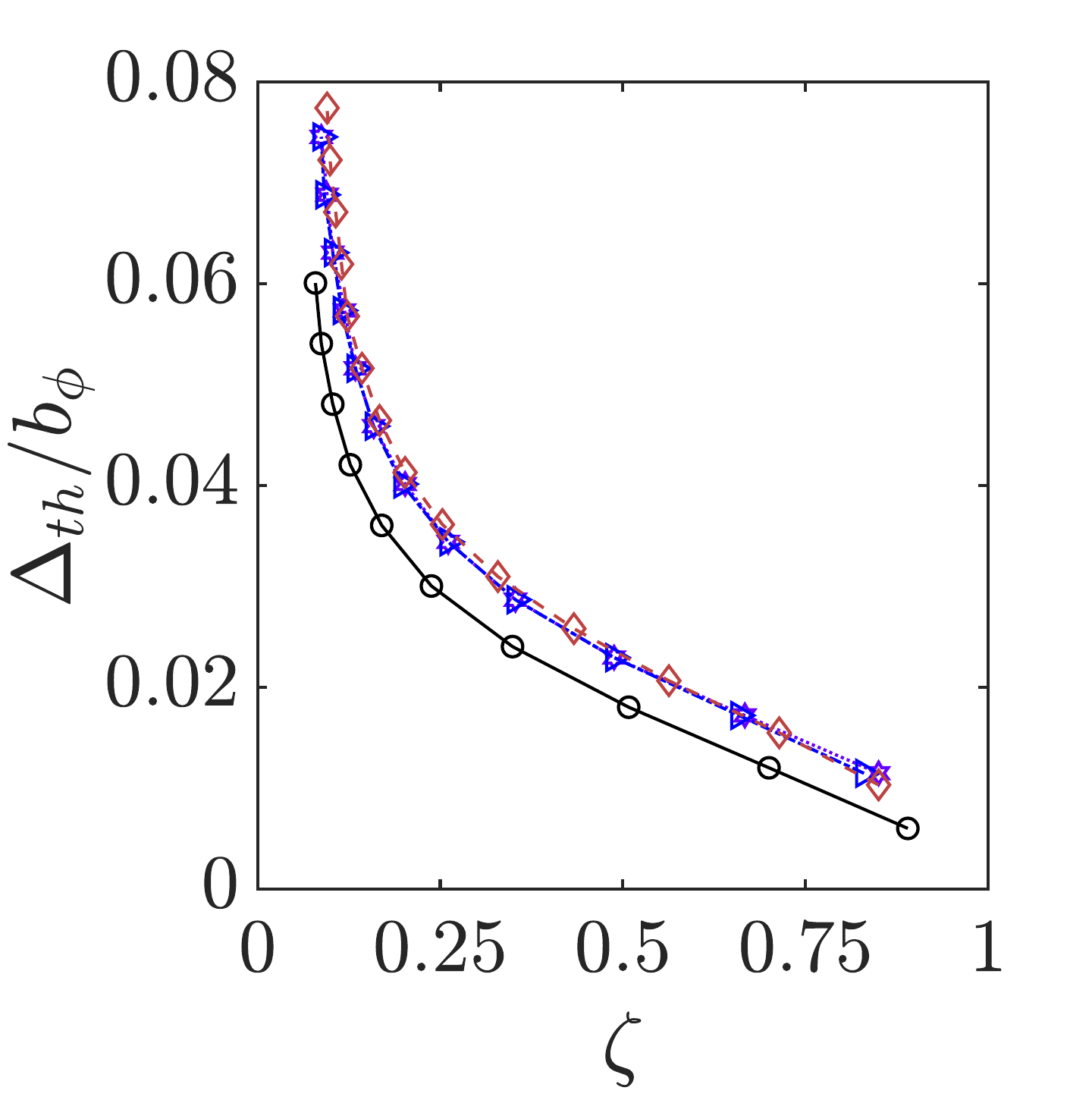}
\caption{(\textit{a}) Conditionally averaged profiles of mean concentration \citep{Kohan&Gaskin2022}. (\textit{b}) Conditional cross correlation function with probe at $x_n/b_{\phi} = 0.025$. (\textit{c,d}) Gradient of the conditional profiles in (\textit{a}) along the interface-normal coordinates.}
  \label{Fig.3}

\end{figure}
\addtocounter{figure}{-1}
\begin{figure} [t!]
	\caption{(Continued from previous page) (\textit{e}) Thickness of the TNTI and TTI ($\Delta_{th}$) as a function of $\zeta$. Note that the abscissas in (\textit{c}) and (\textit{d}) are non-dimensionalized with the jet-exit diameter, $d$, and local scalar half-width, $b_{\phi}$, respectively. The TNTI and TTI outlines ($x_n = 0$) are shown with the vertical dashed black line, while the vertical dashed-dotted line in (\textit{a,d}) denotes the location of the probe for correlation profiles. The top $x$-axis in the inset of ($d$) is normalized by $\eta_q$.}
\end{figure}

Additional insight regarding the concentration transport near the outline may be obtained by considering the two-point scalar correlation in the interface-normal frame of reference. Figure \ref{Fig.3}(\textit{b}) depicts the concentration cross-correlation,
\begin{equation}
C_{\phi\phi}(x_r, \delta) = \frac{\langle \phi'(x_r) \phi'(x_n+\delta) \rangle_I}{\langle \phi^{rms}(x_r)\rangle_I \langle \phi^{rms}(x_r + \delta)\rangle_I},
\label{Eq.2} 
\end{equation}
with the reference point, $x_r$, situated within the finite thickness of the scalar interfacial layer at $x_n/b_{\phi} = 0.025$ for all cases. Note that this point is clearly located within the region characterized by the sharp jump/discontinuity in mean concentration (the dashed-dotted line in figure \ref{Fig.3}\textit{a}). We also checked that small modifications of the reference point location do not alter the results. Here, $\delta$ denotes the lag distance along the interface-normal coordinate, moving into the jet region. The stronger correlation observed between the passive scalar values in the interfacial layer and the core reaffirms the notion of enhanced transport in background turbulence. This is attributed to increased turbulent diffusion and mean radial velocities at the edges of the jet in external turbulence, as shown by the velocity measurements of \cite{Khorsandi2013}.

The thickness of the scalar interfacial layers can be estimated by exploiting the quasi-step jump in the profiles of $\langle \phi \rangle_I$. In particular, the thickness of the scalar TNTI and TTI layers is evaluated by applying a threshold to the gradient of the conditional concentration, $\langle \phi \rangle_I^{'} = \textrm{d}\langle \phi \rangle_I/\textrm{d}x_n$ (figures \ref{Fig.3}\textit{c,d}). The threshold is defined as $\langle \phi \rangle_I^{'} = \zeta [ \langle \phi \rangle_I^{'} ]_{max}$, noting that the results presented herein are largely insensitive to the specific choice of $\zeta$ for $\zeta \in [0.09, 0.9]$.
Subsequently, the extent of the interfacial layer is calculated as the distance between $x_n = 0$ (i.e. the outline) and the intersection of $\langle \phi \rangle_I^{'}/[ \langle \phi \rangle_I^{'} ]_{max}$ and the selected threshold. Figure \ref{Fig.3}(\textit{c}) reveals that the absolute values of the interfacial thickness follow the hierarchy of $\xi$, that is, the adjustment of passive scalar between the jet and the ambient is delayed due to the presence of external turbulence. Figure \ref{Fig.3}(\textit{d}) presents the same data as figure \ref{Fig.3}(\textit{c}), with the difference being that the interface-normal coordinate is normalized by the local concentration half-width; a potentially relevant length scale. It is again evident that background turbulence acts to widen the extent of the scalar interfacial layer at the edges of the jet relative to a non-turbulent ambient. Meanwhile, cases with external turbulence exhibit a reasonable degree of collapse, i.e., the thickness of the scalar TTIs, when normalized by the half-width, remains invariant with background turbulence intensity (more visibly illustrated in figure \ref{Fig.3}\textit{e}). It is also worthwhile mentioning that a value of $\zeta = \mathcal{O}(0.1)$ has been utilized in a number of studies to estimate the thickness of the scalar and vorticity interfacial layers (e.g. \citealp{Wu2019}; \citealp{Jahanbakhshi2021}, among others). The inset of figure \ref{Fig.3}(\textit{d}) shows that upon employing $\zeta = 0.1 \sim 0.2$, a value of $\mathcal{O}(10\eta_q)$ or $\mathcal{O}(\lambda_q)$ is recovered as the thickness of the scalar adjustment region for the cases studied herein, which is in line with previous experiments on the scalar interfacial layer \citep[][]{Gampert2013, Kohan&Gaskin2020}.

The scaling of the TNTI and TTI has important implications in the turbulent entrainment process. For example, \cite{Jahanbakhshi&Madnia2018} demonstrated that a thinner viscous superlayer - which, together with the turbulent sublayer, form the vorticity TNTI - leads to slower propagation of the turbulent front into the surrounding environment. This is a possible explanation for the reduced entrainment rate in a turbulent ambient, as \cite{Kankanwadi&Buxton2022} showed that viscous diffusion plays a much more subdued role in the TTI as compared to inviscid vortex stretching. Consequently, the viscous superlayer in a turbulent ambient should be thinner (if existent at all) compared to that in a quiescent background. In their recent DNS of a mixing layer with $Sc = 1$ in temporally evolving grid turbulence, \cite{Nakamura2023} showed that the scalar and vorticity TTI have comparable thickness. If the scalar TTI were also to follow that of the vorticity for $Sc \gg 1$
, the results of figure \ref{Fig.3}(\textit{c,d}) would suggest a significant increase in the thickness of the turbulent sublayer to offset the thinner/absent viscous superlayer in external turbulence. In any case, whilst the results of figure \ref{Fig.3} pertain to the scalar TNTI and TTI, they could be useful to further broaden our understanding of the entrainment phenomenon.

\subsection{Entrainment/detrainment analysis \label{Entrainment analysis}}

The concentration holes (also referred to as bubbles) represent regions with $\phi < \phi_t$ that lie within the jet. Only holes with an area, $A_h$, greater than 4 $\textrm{pixels}^2$ (approximately 6.7 $\eta_q^2$ in the physical world) are considered hereafter, to account for experimental noise \citep{Kohan&Gaskin2023}. The origin of the holes in a turbulent flow is either (i) within the flow itself \citep{daSilva2014} due to the internal intermittency of the velocity and scalar field or (ii) due to the large-scale entrainment (engulfment) of the ambient fluid. Without a Lagrangian approach, it is difficult to determine whether the scalar holes originate from the internal mechanics of the turbulence or are drawn into the primary shear flow from the surrounding environment. A methodology akin to that of \cite{Xu2023} is therefore adopted to distinguish the engulfed holes from those generated within the shear flow. Specifically, the concentration homogeneity within the scalar holes is investigated as a function of the Euclidean distance of their centroids to the interface, $d_h$. The r.m.s. concentration within the hole is defined as
\begin{equation}
\phi^{rms}_h = \sqrt{\overline{\left( \phi - \phi_h \right)^2}},
\label{Eq.3}
\end{equation}

\begin{figure}
\centering
\raisebox{1.75in}{(\textit{a})}\includegraphics[width = 0.328\textwidth]{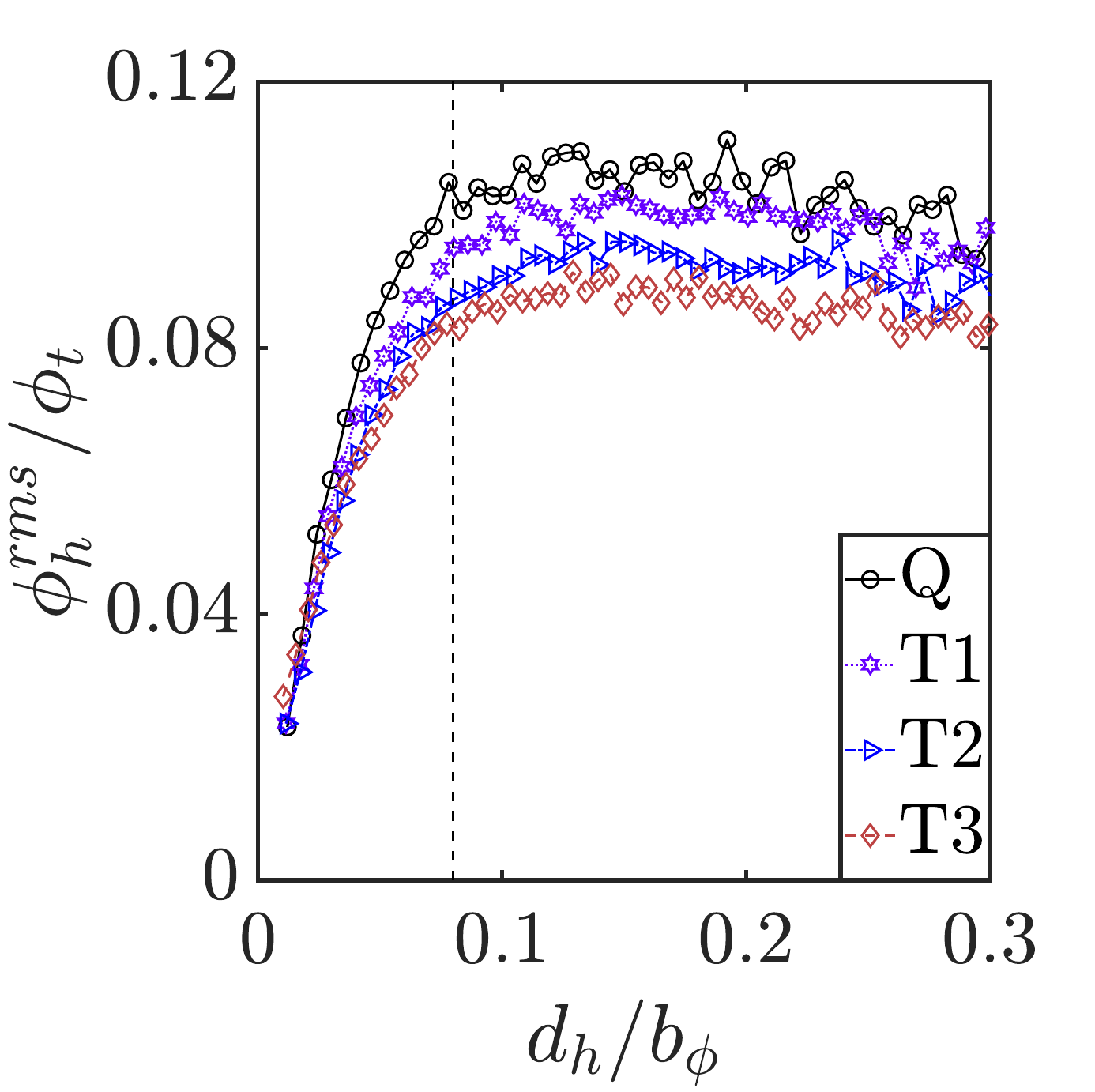}\hspace*{-0.3cm}
\raisebox{1.75in}{(\textit{b})}\includegraphics[width = 0.328\textwidth]{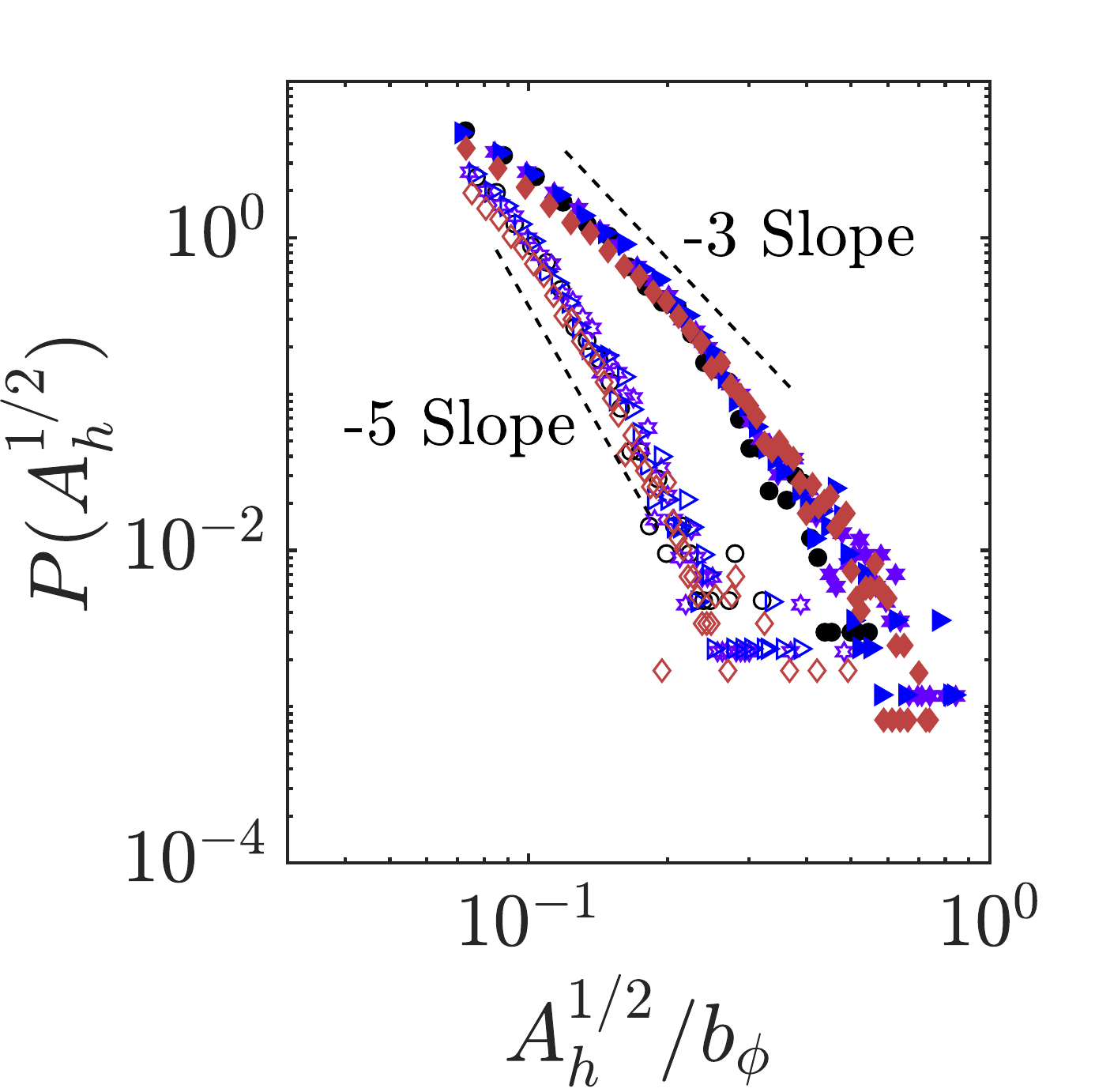}\hspace*{-0.3cm}
\raisebox{1.75in}{(\textit{c})}\includegraphics[width = 0.328\textwidth]{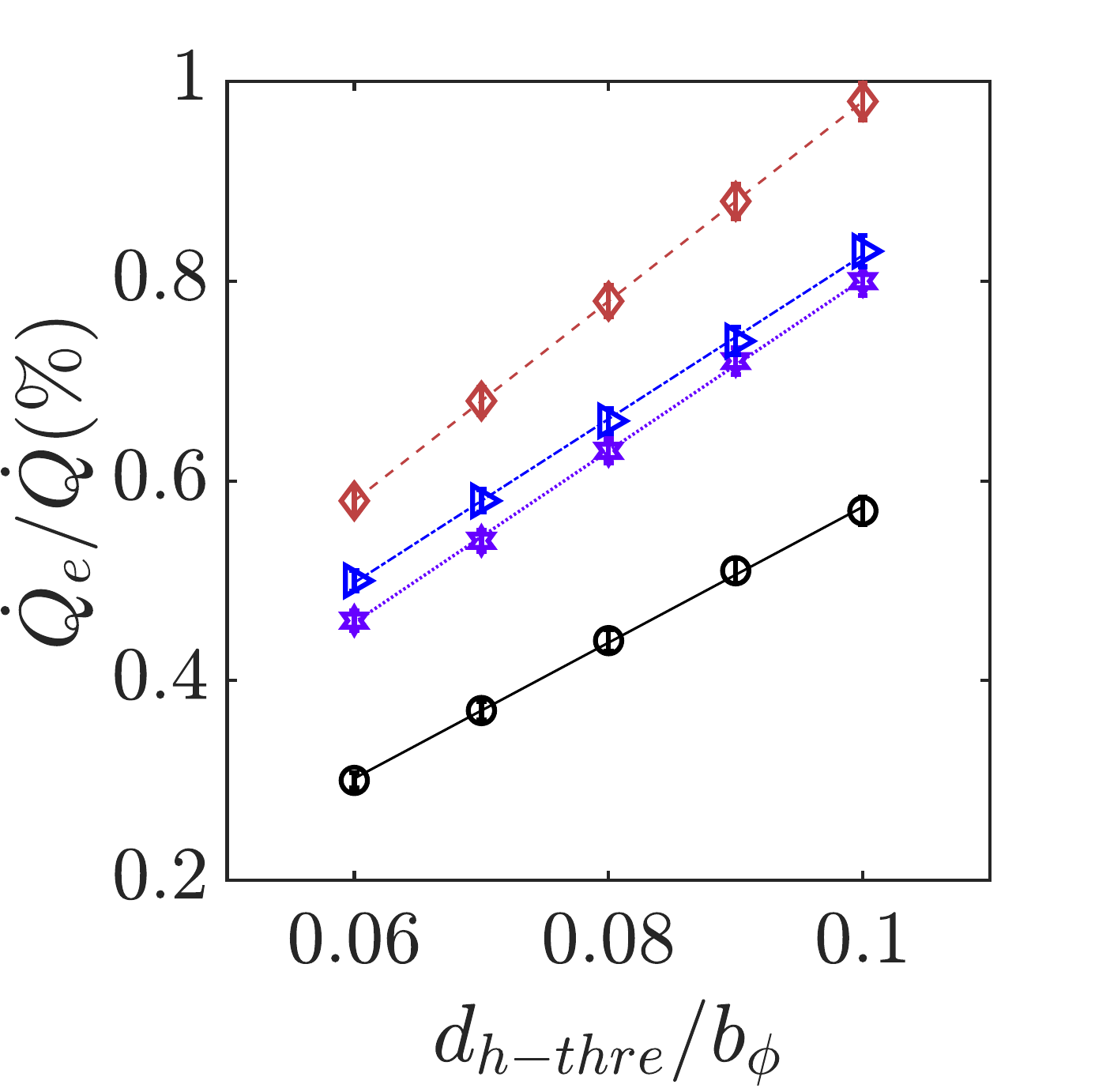}
\caption{(\textit{a}) Conditionally averaged profiles of r.m.s. concentration within the holes, normalized by the scalar thresholds that detect the interfaces. (\textit{b}) P.d.f. of the characteristic length of the holes within the interfacial layer ($d_h < 0.08 b_{\phi}$, hollow markers) and inside the turbulent region ($d_h > 0.08 b_{\phi}$, filled markers). (\textit{c}) Sensitivity of the assessed engulfed flux against the demarcation line, $d_{h-thre}$. The lines and error bars in (\textit{c}) denote the best linear regression fit and the 95\% confidence interval, respectively.}
\label{Fig.4}
\end{figure}

\noindent where $\phi_h$ represents the mean concentration within the hole. The ambient and the jet regions are characterized by low and high levels of concentration fluctuations, respectively. Hence, it is reasonable to anticipate that the engulfed holes, which are `trapped' in the jet by large-scale events near the interface, possess relatively low $\phi^{rms}_h$. Figure \ref{Fig.4}(\textit{a}) presents the variation of $\phi^{rms}_h$ with distance from the TNTI and TTI outlines. Analogous to the velocity field \citep{Xu2023}, the scalar field inside the holes is initially uniform and gradually becomes inhomogeneous as the hole is positioned deeper into the shear flow. This trend is captured in figure \ref{Fig.4}(\textit{a}), in which the scalar holes are divided into two regions with $d_{h-thre} \approx 0.08 b_{\phi}$ as the border for all cases. Interestingly, $0.08 b_{\phi}$ is very close to the thickness of the scalar interfacial layer for values of $\zeta = \mathcal{O}(0.1)$ (figure \ref{Fig.3}\textit{e}), indicating that the behavior of holes within the interfacial layer is essentially different from those in the turbulent core region \citep{Jahanbakhshi&Madnia2016, Long2022}. The probability density function (p.d.f.) of the characteristic length of the holes, $A_h^{1/2}$, in the two regions is displayed in figure \ref{Fig.4}(\textit{b}), where it is found that the p.d.f.s inside the turbulent core (filled markers) appear to follow the $-3$ power-law. This scaling was previously reported by \cite{daSilva2014} in DNS of homogeneous isotropic turbulence (HIT), in which all the holes originate from the internal turbulence. We thus infer that holes with $d_h > 0.08 b_{\phi}$ are statistically similar to those in HIT, and most likely not related to engulfment. On the other hand, the distribution of the p.d.f.s within the interfacial layer (hollow markers) is closer to a $-5$ power-law. This result falls between the $-4$ power-law distribution in a Mach-0.2 shear layer \citep{Jahanbakhshi&Madnia2016} and the $-6$ power-law distribution in an incompressible jet \citep{Xu2023} observed for the p.d.f. of the characteristic size of the holes within the interfacial layer. This evidence suggests that holes with $d_h < 0.08b_{\phi}$ are likely engulfed by the energetic large-scale motions of the jet. Lastly, we note that the analysis of the isolated ambient holes from our two-dimensional (2-D) dataset can be subjected to interpretation artefacts. For example, some of the detected holes might be in fact low-concentration regions directly connected to the ambient, termed `pockets'. Nevertheless, the findings of figure \ref{Fig.4}(\textit{b}) reveal that the vast majority of the identified holes appear to be isolated patches rather than pockets, since the p.d.f. of the characteristic size of the pockets, as viewed in 2-D slices, is governed by a milder power-law, closer to $-1$, as recently found empirically in DNS of turbulent boundary layers \citep{Long2022}. Furthermore, the p.d.f.s exhibit excellent collapse, indicating that the experimental constraints are not likely to bias the results across different background conditions. These reasons lend credibility to the engulfment analysis carried out below.

Table \ref{Tab.1} shows that ambient turbulence increases the relative area of the holes within the jet, despite a larger jet area. However, this does not necessarily indicate that engulfment is increased in the turbulent ambient, as only holes with $d_h < 0.08b_{\phi}$ are related to the engulfment, per the discussion above. The contribution of the engulfed mass flux to the total flux is estimated using \citep{Westerweel2009},
\begin{equation}
\frac{\dot{Q}_e}{\dot{Q}} \approx \frac{2\pi \rho u_c \int_{0}^{\infty} \mathcal{P}_e(r)\, r\, \textrm{d}r}{2\pi \rho u_c \int_{0}^{\infty} \mathcal{P}(r)\, r\, \textrm{d}r},
\label{Eq.4}
\end{equation}
where, $\rho$, $u_c$, and $r = (y^2 + z^2)^{1/2}$ denote the density, ensemble-averaged jet centerline velocity, and radial coordinate, respectively. Furthermore, $\mathcal{P} (r)$ represents the probability for a given fluid element in location $r$ to be bounded by the outline, while $\mathcal{P}_e(r)$ is the probability for a given element to be a hole whose distance to the interface is \emph{less} than $d_{h-thre}$. The ratio $\dot{Q}_e/\dot{Q}$ amounts to (0.44 $\pm$ 0.010)\%, (0.63 $\pm$ 0.012)\%, (0.66 $\pm$ 0.013)\%, and (0.78 $\pm$ 0.015)\% for cases Q, T1, T2, and T3, respectively, using $d_{h-thre} = 0.08 b_{\phi}$. This result indicates that the contribution of engulfment is indeed increased in background turbulence, albeit slightly. Nibbling, or a third mechanism such as increased turbulent diffusion, appear to remain the dominant entrainment process in the far field of a jet, at least with the moderate range of background turbulence intensities and length scales investigated herein ($0.15 \leqslant \xi \leqslant 0.26$ and $3.5 \leqslant \mathcal{L} \leqslant 5.5$).
The current value of $\dot{Q}_e/\dot{Q}$ in the quiescent ambient is lower than the reported value of $6\%-8\%$ in scalar experiments of jets and plumes in a quiescent ambient (e.g. \citealp{Westerweel2009}; \citealp{Parker2019}). This difference may be ascribed to the higher Reynolds number used here (e.g., $Re_J = 10600$ compared to $Re_J = 2000$ in \cite{Westerweel2009}), surpassing the so called `transition' Reynolds number of $Re_J \approx 10^4$ \citep{Dimotakis2000}. It is argued that the level of scalar unmixedness greatly reduces beyond this transition value, thereby reducing the area of the scalar holes and the perceived importance of engulfment. Moreover, the abovementioned studies use all the ambient holes encompassed by the interface `envelope' in equation \ref{Eq.4}, rather than the actual convoluted outline, thus obtaining an upper limit for the contribution of engulfment.
Using Lagrangian statistics, \cite{Taveira2013} also showed that the contribution of engulfment to the total entrainment rate is less than 1\% for temporal planar jets in a quiescent ambient. The robustness of the engulfment flux estimation is illustrated in figure \ref{Fig.4}(\textit{c}), where it is seen that variations of $\pm 25\%$ in $d_{h-thre}$ do not affect the aforementioned conclusion, that is, external turbulence promotes the large-scale entrainment in a jet.

\begin{table}
 \begin{center}
 \def~{\hphantom{0}}
     \hrule
		\begin{tabular}{l c c c}
			Case & $A_j/d^2\, (A_j/b_{\phi}^2)$ & $A_h/A_j$ (\%) & $A_i/A_j$ (\%)  \Tstrut\Bstrut \\
            \tikz\draw[thick, color = black] (0,0) circle (0.6ex); Q & 51.6 (8.1) & 1.55 & 4.36\\
            \hexagram{mypurple} T1 & 59.8 (8.8) & 1.99 & 8.53\\
            \emptytriangle{myblue} T2 & 61.8 (8.9) & 2.06 & 9.21\\
            \emptydiamond{myred} T3 & 88.1 (10.2) & 2.78 & 11.85
		\end{tabular}
  \caption{The averaged ratio of holes' and islands' area to that of the main jet for different cases. The averaged area of the jet ($A_j$) is also provided. Note that $A_j$ is calculated as the area enclosed by the TNTI and TTI outlines, including the holes. The width of the 95\% confidence interval for all the reported parameters is within 5\% of their respective mean values.}
 \hrule
		\label{Tab.1}
	\end{center}
\end{table}

It is well understood that the large vorticity structures are responsible for inviscid engulfment of ambient fluid at the inward cusps of the turbulence/ambient interface (see e.g. \citealp{Yule1978}; \citealp{Ferre1989}; \citealp{Philip&Marusic2012}). Engulfment in jets can be envisioned as a two-part process, comprising an induced inflow towards the interface and the subsequent entrapment of ambient fluid. Turbulence in the background tends to disrupt the large-scale coherent structures of the jet \citep{Hunt1994}, reducing the streamwise velocity and hindering the generation of the entrainment wind as compared to the non-turbulent ambient \citep{Khorsandi2013}. We, however, postulate that the increased large-/small-scale indentations of the interface in the presence of external turbulence \citep{Kohan&Gaskin2022} aid the entrapment of ambient fluid parcels, leading to an overall greater engulfment, despite the reduced entrainment wind. The result of increased engulfment in the turbulent ambient mirrors that of \cite{Kankanwadi&Buxton2023}, who hypothesized that free-stream turbulence acts to enhance the large-scale entrainment in cylindrical wakes.

Islands are defined as patches of scalar with $\phi \geqslant \phi_t$, which appear disconnected from the main jet region in our planar measurements. Small air bubbles occasionally enter the FOV, causing non-physical large PLIF signals, and, hence, concentration values. Islands with a mean concentration greater than $3.5 \phi_t$ are therefore discarded to address this problem. A sensitivity analysis was performed to ascertain that variations within the neighbourhood of the chosen coefficient of 3.5 did not impact the average area of the islands. Table \ref{Tab.1} reveals that external turbulence result in an increased contribution of the islands' area to that of the jet, noting that the ratio $A_i/A_j$ is arguably underestimated in the turbulent ambient due to the extensive range of the islands and the limited size of the FOV (see, e.g., figure \ref{Fig.1}\textit{b}) (Multimedia available online). An increased occurrence of detached islands can be cautiously interpreted as strong local detrainment events, since the vortical structures containing the passive scalar are breaking away from the interface. However, the same experimental limitations that complicated the detection of the isolated holes further complicate the identification of the genuine detrainment patches. Essentially, the detected islands may be attached to the jet region in a different streamwise plane through three-dimensional re-connection events, e.g., in the form of a `tea-cup handle' \citep{Borrell&Jimenez2016}. This is particularly an issue in the turbulent background due to the increased topological irregularities of the interface. We, therefore, use experimental visualizations to aid our narrative.

\begin{figure}
\hspace*{-3.1cm}
\includegraphics[width = 0.75\textwidth]{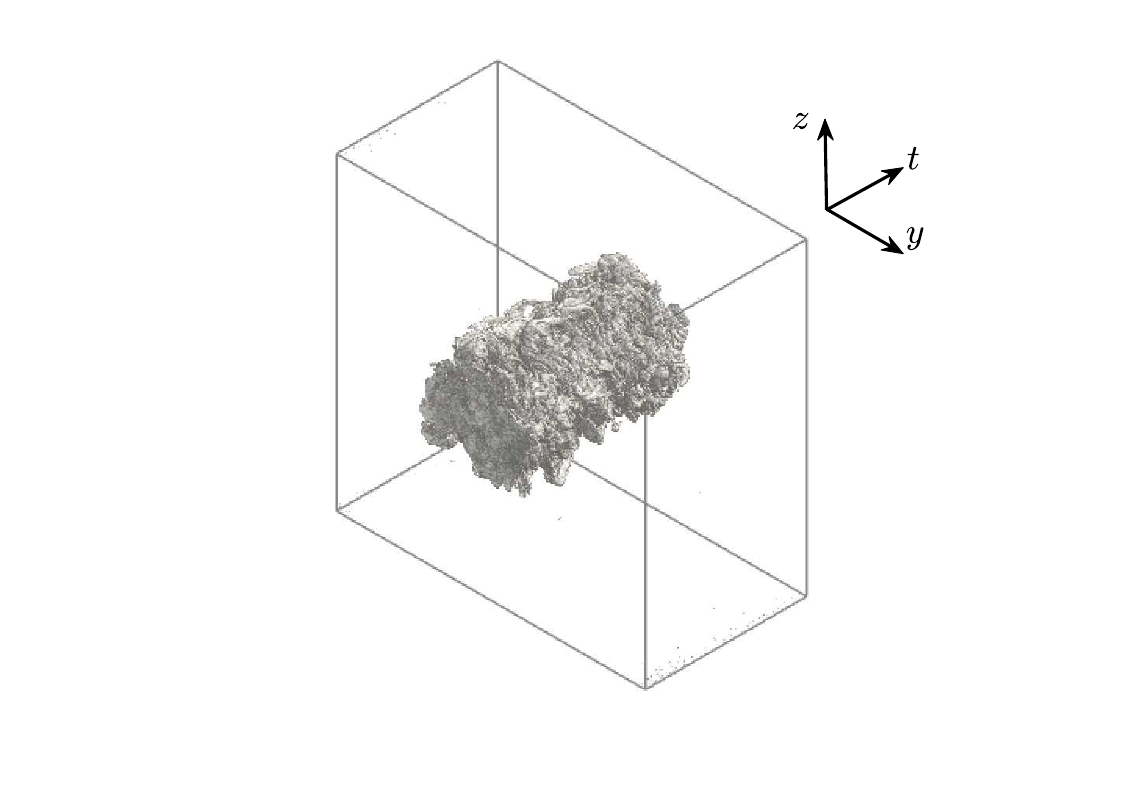}\hspace*{-2.2cm}
\includegraphics[width = 0.75\textwidth]{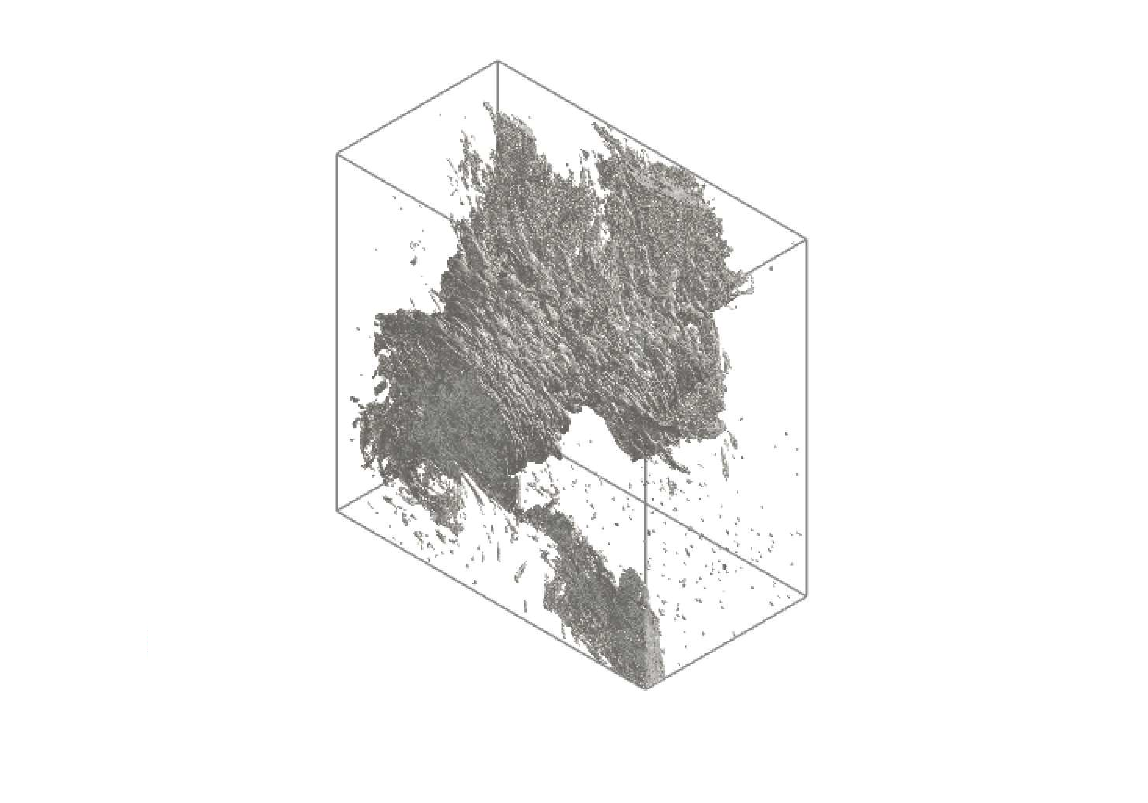}
	\caption{(2+1)-D space-time visualization of isosurface of $\phi_t$ for the jet in case Q (left panel) and case T3 (right panel). The cubes represent the spatial extent of the FOV.}
	\label{Fig.5}
\end{figure}

In a quiescent ambient, the islands are usually re-entrained into free-shear flows \citep{Hussain&Clark1981} and boundary layers \citep{Wu2020} within a few eddy turnover times. This is evident in the video of figure \ref{Fig.1}(\textit{a}) (Multimedia available online), as the islands in the quiescent ambient are generally close to the TNTI outline. In a turbulent ambient, however, the islands are located further away from the jet and diffuse faster into the background (as seen in the experiments, the video of figure \ref{Fig.1}\textit{b}) (Multimedia available online), hindering the possibility of re-entrainment. Note that the small `gap' in the video of figure \ref{Fig.1}(\textit{b}) (Multimedia available online) corresponds to invalid snapshots, where the FOV cannot contain the full extent of the TTI outline \citep{Kohan&Gaskin2022}. Figure \ref{Fig.5} shows (2+1)-D space-time visualizations of the jet for cases Q and T3 at the isocontours defined by the concentration threshold, $\phi_t$, corresponding to the videos of figure \ref{Fig.1} (Multimedia available online). This has been achieved by stacking 526 contiguous PLIF snapshots, similar to \cite{Shan&Dimotakis2006}. It is worth noting that the evolution of the jet with downstream distance is not captured in figure \ref{Fig.5}, as the third dimension is essentially time. The present space-time data are, nonetheless, valuable for our purposes as they clearly demonstrate the large- and small-scale undulations of the interface and the intense detrainment events in the presence of external turbulence.

Compared to a non-turbulent ambient, the jet in external turbulence detrains more frequently due to the competition between the jet and background turbulence to entrain fluid from one another. When the TTI outline propagates towards the jet core (i.e. detrainment), large patches of passive scalar are left in the ambient. In the absence of a strong entrainment wind, these patches disperse from the interface and diffuse in the ambient by the action of turbulent eddies. The latter aligns with the concept of increased effective eddy diffusivity in the turbulent background, and suggests an increase in local detrainment events as compared to the quiescent ambient. This description also supports the hypothesis of \cite{Westerweel2009} regarding increased detrainment in external forcing and also the findings of \cite{Kankanwadi&Buxton2020}, who noticed extreme detrainment events for a wake in free-stream turbulence.

\begin{figure}
\centering
\raisebox{1.78in}{(\textit{a})}\includegraphics[width = 0.43\textwidth]{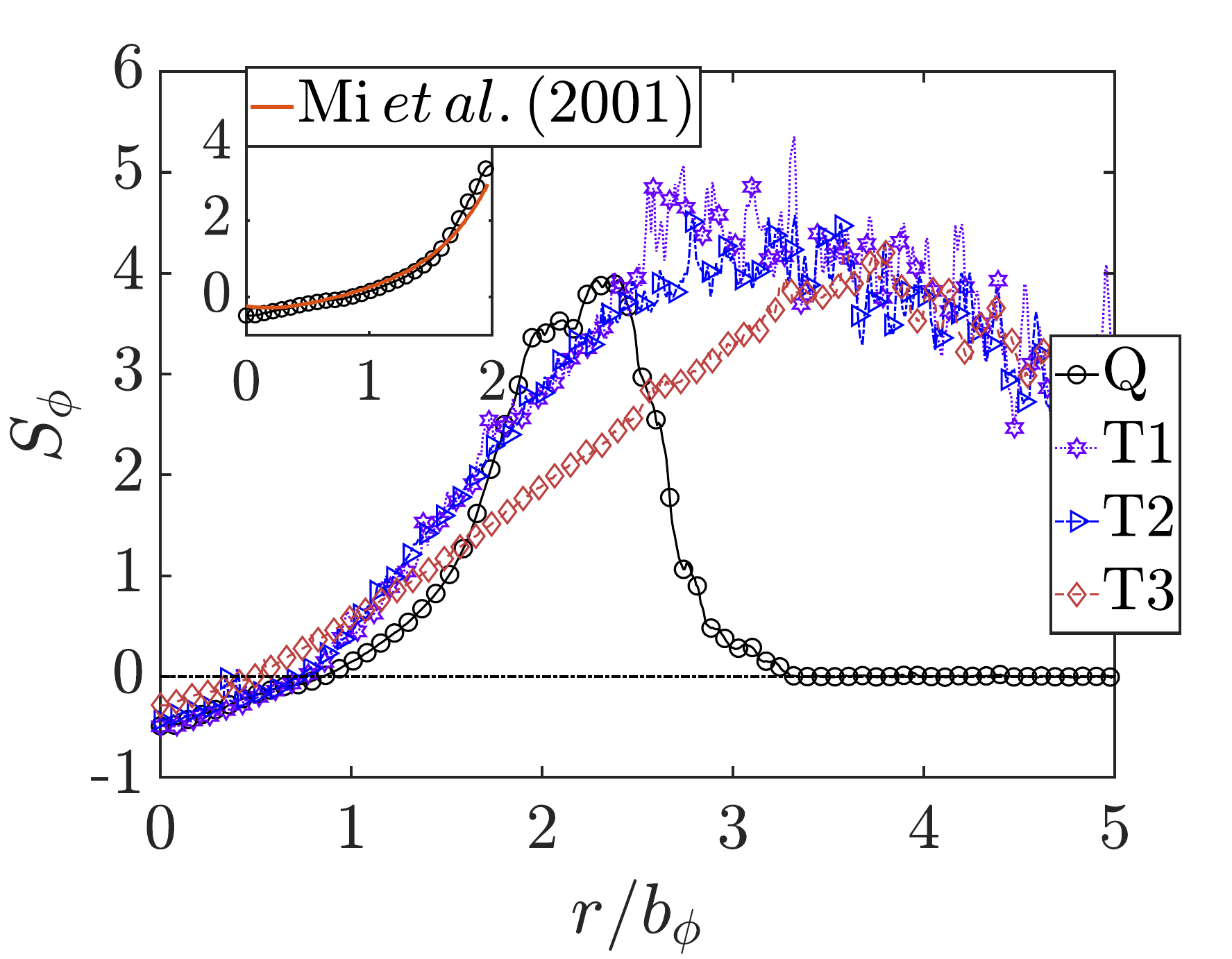}
\raisebox{1.78in}{(\textit{b})}\includegraphics[width = 0.43\textwidth]{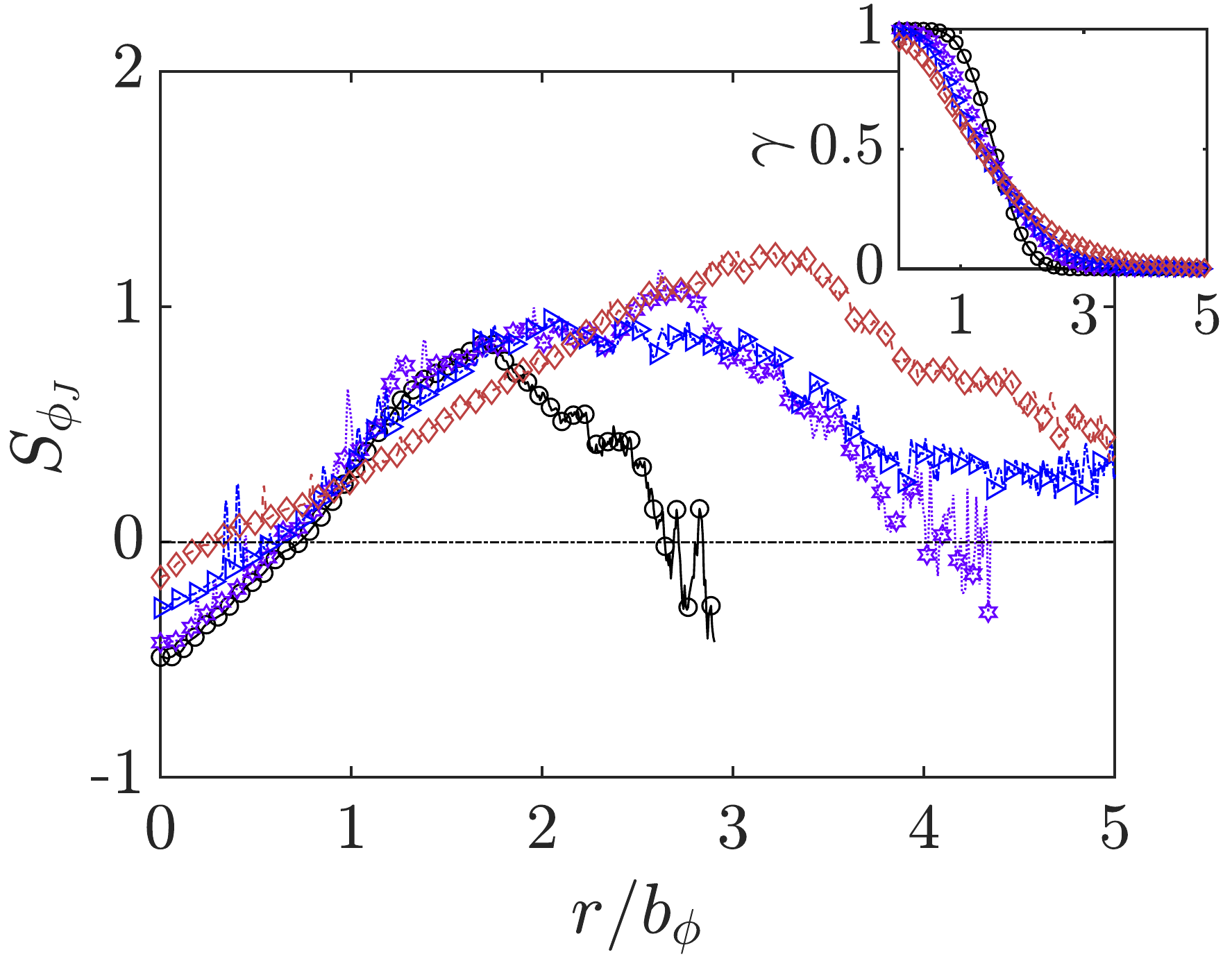}
\caption{(\textit{a}) Radial profiles of the scalar skewness for the studied cases. The inset displays $S_{\phi}$ in the quiescent ambient and that of \cite{Mi2001} at $x/d = 25$. (\textit{b}) Zonal average of the scalar skewness within the jet. The inset shows the effect of external turbulence on the intermittency factor, $\gamma$.}
\label{Fig.6}
\end{figure}

Consistent with the above description, \citet{Khorsandi2013} showed that external turbulence generated by the RJA tends to lower the entrainment into the far field of the jet. This notion is corroborated by investigating the radial profiles of the concentration skewness, defined as
\begin{equation}
S_{\phi} = \frac{ \overline{ \left( \phi - \overline{\phi} \right)^3}}{\left[ \overline{ \left( \phi - \overline{\phi} \right)^2} \right]^{3/2}}.
\label{Eq.5}
\end{equation}
\noindent In the context of scalar entrainment, positive $S_{\phi}$ implies that fluid parcels containing the passive scalar are being mixed in a background of un-dyed fluid (i.e. $\phi > \overline{\phi}$ on average), whereas negative skewness suggests mixing of low-concentration patches within the shear flow, that is, $\phi < \overline{\phi}$ on average. Therefore, $S_{\phi}$ can act as a suitable surrogate for net entrainment into the jet. Figure \ref{Fig.6} shows the effect of ambient turbulence on the scalar skewness of an axisymmetric jet. Prior to interpreting the results, we note that the skewness profile in the quiescent ambient shows good agreement with temperature measurements of \cite{Mi2001}, indicating that the current experimental resolution accurately captures the third order scalar statistics. For a non-turbulent ambient, the value of $S_{\phi}$ is always negative at the center of fully developed shear flows \citep[e.g.][]{Pope2000} due to the entrainment of low-concentration background parcels. This trend also persists for the jet in our turbulent ambient cases (figure \ref{Fig.6}\textit{a}) but, more importantly, the less negative values of $S_{\phi}$ within the core ($r/b_{\phi} \lesssim 1$) hint at the lowered entrainment into the jet. It is also worth mentioning that strong but sporadic meandering events in the turbulent ambient can even displace the jet center in space, causing artificially negative $S_{\phi}$. This in turn leads to the overestimation of the entrained ambient parcels as inferred from the skewness profiles. This is addressed in figure \ref{Fig.6}(\textit{b}), where the zonal average of scalar skewness within the jet, $S_{\phi_J} = \overline{(\phi - \overline{\phi_J})^3_J}/[\overline{(\phi - \overline{\phi_J})^2_J}]^{3/2}$, is presented along with the intermittency factor, $\gamma$, to show the effect of external intermittency due to meandering. The zonal average of quantity $\mathcal{C}$ is defined as $\overline{\mathcal{C}_J} = \overline{I\mathcal{C}}/\overline{I}$, where $I$ is the intermittency function, set to zero in the ambient and to unity in the jet, so that $\gamma = \overline{I}$. It is appreciated that the values of $S_{\phi_J}$ remain closer to the Gaussian value of 0, owing to the removal of external intermittency by zonal sampling. Compared to $S_{\phi}$, the profiles of $S_{\phi_J}$ in the turbulent ambient deviate further from their quiescent counterpart for $r/b_{\phi} \lesssim 1$, and better portray the effect of external forcing on the distribution of the scalar field within the jet. In summary and in accordance with the increased detrainment events, the results of figure \ref{Fig.6} imply that the net entrainment of ambient fluid is suppressed in the presence of external turbulence. 

\section{Conclusions \label{Conclusions}}
Planar laser-induced fluorescence measurements were conducted to investigate the effect of zero-mean-flow approximately homogeneous background turbulence on the outer region of an axisymmetric jet. The external turbulence is generated using a random jet array.

\begin{figure}
\centering
	{\includegraphics[width = 0.82\textwidth]{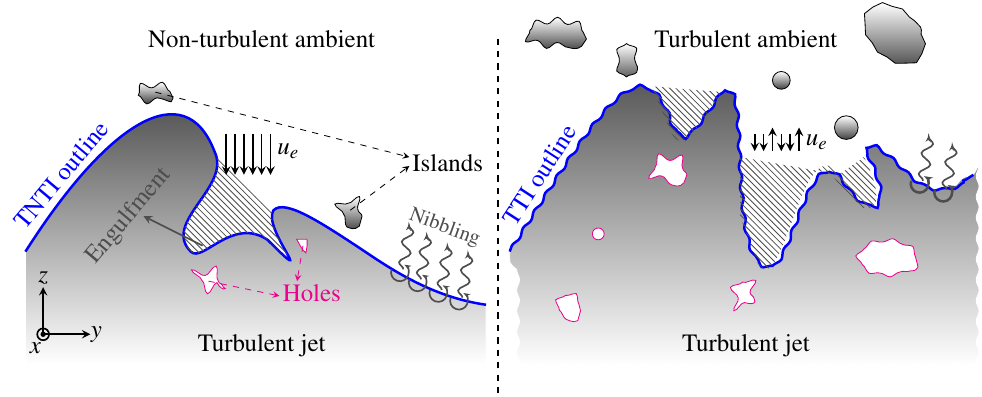}}
	\caption{Conceptual model of the behavior of the outer jet region in the presence of ambient turbulence. Note that $u_e$ denotes the entrainment wind, induced by the large-scale motions of the jet. The hatched regions represent entrapment (engulfment) of ambient fluid.}
	\label{Fig.7}
\end{figure}

Figure \ref{Fig.7} summarizes the effect of ambient turbulence on the interfacial properties and entrainment into an axisymmetric jet. The action of external turbulence is to increase the large- and small-scale modulations of the interface \citep{Kohan&Gaskin2022} and to increase the thickness of the scalar interfacial layer. The two-point statistics revealed enhanced scalar transport towards the edges of the jet in the turbulent ambient, owing to increased turbulent diffusion and mean radial velocities \citep{Khorsandi2013}. Whilst external turbulence reduces the magnitude of the entrainment wind, $u_e$, by disrupting the large eddying motions of the jet \citep{Hunt1994}, the contribution of engulfment to the total mass flow rate is slightly enhanced (although still below 1\%), seen as an increased presence of ambient holes within the finite thickness of the interfacial layer. This phenomenon can be potentially explained as the increased entrapment of low-concentrated ambient fluid elements between the inward cusps of the TTI outline.

The current planar visualizations showed that turbulence in the ambient results in an increased occurrence of concentration islands far away from the jet outline. This hints at local detrainment episodes due to the competition between the turbulent ambient and the jet to entrain fluid from one another, and can result in a reduced entrainment rate into the jet, despite a larger interfacial surface area. In accordance with the previous observation, the radial scalar skewness profiles also elucidated that mixing of ambient fluid inside the jet is enhanced in the turbulent ambient. The present qualitative experimental approach pertaining the detrained patches may shed light on the detrainment mechanism in other turbulent flows and for other background conditions.

Lastly, in \cite{Kohan&Gaskin2022} we cautiously alluded to the possible universality of the impact of external turbulence on the topology of the TTI outline in jets and wakes. One can hereby extend this hypothesis to the entrainment mechanisms of jets and wakes. This extension is supported by our findings in the context of a jet subjected to turbulence generated by a RJA, which align well with those in a wake subjected to grid turbulence, particularly with regard to intense detraining events \citep{Kankanwadi&Buxton2020} and the increased significance of engulfment \citep{Kankanwadi&Buxton2023} in the presence of external forcing.

\subsubsection*{Acknowledgements}
The research was supported by funding provided by the Natural Sciences and Engineering Research Council of Canada discovery grants (RGPIN 2016-04473 and RGPIN 2022-03438). K. F. Kohan also acknowledges the Mitacs Globalink Research Award (FR109660).

\subsubsection*{Author declarations}
Conflict of interest: The authors have no conflicts to disclose.
\\Data Availability: The data that support the findings of this study are available from the corresponding author upon reasonable request.

\subsubsection*{Author ORCID}
K. F. Kohan, \url{https://orcid.org/0000-0002-2965-7992};\\ S. J. Gaskin, \url{https://orcid.org/0000-0003-2036-2124}.

\bibliography{Kohan}{}

\begin{thebibliography}{45}
\expandafter\ifx\csname natexlab\endcsname\relax\def\natexlab#1{#1}\fi
\def\au#1{#1} \def\ed#1{#1} \def\yr#1{#1}\def\at#1{#1}\def\jt#1{\textit{#1}} \def\bt#1{#1}\def\bvol#1{\textbf{#1}} \def\vol#1{#1} \def\pg#1{#1} \def\publ#1{#1}\def\arxiv#1{#1}\def\org#1{#1}\def\st#1{\textit{#1}}

\bibitem[Bisset {\em et~al.\/}(2002)Bisset, Hunt \& Rogers]{Bisset2002}
{\sc \au{Bisset, D.~K.}, \au{Hunt, J. C.~R.} \& \au{Rogers, M.~M.}} \yr{2002}  \at{The turbulent/non-turbulent interface bounding a far wake}.  \jt{J. Fluid Mech.}  \bvol{451},  \pg{383--410}.

\bibitem[Bonnebaigt {\em et~al.\/}(2016)Bonnebaigt, Caulfield \& Linden]{Bonnebaigt2016}
{\sc \au{Bonnebaigt, R.}, \au{Caulfield, C.~O.} \& \au{Linden, P.~F.}} \yr{2016}  \at{Detrainment of plumes from vertically distributed sources}.  \jt{Environ. Fluid Mech.}  \bvol{18},  \pg{3--25}.

\bibitem[Borrell \& Jim\'enez(2016)]{Borrell&Jimenez2016}
{\sc \au{Borrell, G.} \& \au{Jim\'enez, J.}} \yr{2016}  \at{Properties of the turbulent/non-turbulent interface in boundary layers}.  \jt{J. Fluid Mech.}  \bvol{801},  \pg{554--596}.

\bibitem[Chen \& Buxton(2023)]{Chen&Buxton2023}
{\sc \au{Chen, J.} \& \au{Buxton, O. R.~H.}} \yr{2023}  \at{Spatial evolution of the turbulent/turbulent interface geometry in a cylinder wake}.  \jt{J. Fluid Mech.}  \bvol{969},  \pg{A4}.

\bibitem[Dimotakis(2000)]{Dimotakis2000}
{\sc \au{Dimotakis, P.~E.}} \yr{2000}  \at{The mixing transition in turbulent flows}.  \jt{J. Fluid Mech.}  \bvol{409},  \pg{69--98}.

\bibitem[Eames \& Fl\'or(2022)]{Eames&Flor2022}
{\sc \au{Eames, I.} \& \au{Fl\'or, J-B.}} \yr{2022}  \at{Spread of infectious agents through the air in complex spaces}.  \jt{Interface Focus}  \bvol{12 \textnormal{(10)}},  \pg{20210080}.

\bibitem[Ferre {\em et~al.\/}(1989)Ferre, Mumford, Savill \& Giralt]{Ferre1989}
{\sc \au{Ferre, J.~A.}, \au{Mumford, J.~C.}, \au{Savill, A.~M.} \& \au{Giralt, F.}} \yr{1989}  \at{Three-dimensional large-eddy motions and fine-scale activity in a plane turbulent wake}.  \jt{J. Fluid Mech.}  \bvol{210},  \pg{371--414}.

\bibitem[Friehe {\em et~al.\/}(1971)Friehe, van Atta \& Gibson]{Friehe1971}
{\sc \au{Friehe, C.~A.}, \au{van Atta, C.~W.} \& \au{Gibson, C.~H.}} \yr{1971}  \at{Jet turbulence: dissipation rate measurements and correlations}.  \jt{AGARD Turbul. Shear Flows}  \bvol{18},  \pg{1--7}.

\bibitem[Gampert {\em et~al.\/}(2013)Gampert, Narayanaswamy, Schaefer \& Peters]{Gampert2013}
{\sc \au{Gampert, M.}, \au{Narayanaswamy, V.}, \au{Schaefer, P.} \& \au{Peters, N.}} \yr{2013}  \at{Conditional statistics of the turbulent/non-turbulent interface in a jet flow}.  \jt{J. Fluid Mech.}  \bvol{731},  \pg{615--638}.

\bibitem[Gaskin {\em et~al.\/}(2004)Gaskin, McKernan \& Xue]{Gaskin2004}
{\sc \au{Gaskin, S.~J.}, \au{McKernan, M.} \& \au{Xue, F.}} \yr{2004}  \at{The effect of background turbulence on jet entrainment: an experimental study of a plane jet in a shallow coflow}.  \jt{J. Hydraul. Res.}  \bvol{42 \textnormal(5)},  \pg{533--542}.

\bibitem[Gladstone \& Woods(2014)]{Gladstone&Woods2014}
{\sc \au{Gladstone, C.} \& \au{Woods, A.~W.}} \yr{2014}  \at{Detrainment from a turbulent plume produced by a vertical line source of buoyancy in a confined, ventilated space}.  \jt{J. Fluid Mech.}  \bvol{742},  \pg{35--49}.

\bibitem[Hancock \& Bradshaw(1989)]{Hancock&Bradshaw1989}
{\sc \au{Hancock, P.~E.} \& \au{Bradshaw, P.}} \yr{1989}  \at{Turbulence structure of a boundary layer beneath a turbulent free stream}.  \jt{J. Fluid Mech.}  \bvol{205},  \pg{45--76}.

\bibitem[Hunt(1994)]{Hunt1994}
{\sc \au{Hunt, J. C.~R.}} \yr{1994} Atmospheric jets and plumes.  \bt{In {\em Recent Research Advances in the Fluid Mechanics of Turbulent Jets and Plumes\/} (ed. \ed{P.~A. Davies \& M.~J.~Valente Neves})},  \pg{pp. 309--334}.  \publ{Springer}.

\bibitem[Hunt {\em et~al.\/}(2006)Hunt, Eames \& Westerweel]{Hunt2006}
{\sc \au{Hunt, J. C.~R.}, \au{Eames, I.} \& \au{Westerweel, J.}} \yr{2006}  \at{Mechanics of inhomogeneous turbulence and interfacial layers}.  \jt{J. Fluid Mech.}  \bvol{554},  \pg{499--519}.

\bibitem[Hussain \& Clark(1981)]{Hussain&Clark1981}
{\sc \au{Hussain, A. K. M.~F.} \& \au{Clark, A.~R.}} \yr{1981}  \at{On the coherent structure of the axisymmetric mixing layer: a flow-visualization study}.  \jt{J. Fluid Mech.}  \bvol{104},  \pg{263--294}.

\bibitem[Jahanbakhshi(2021)]{Jahanbakhshi2021}
{\sc \au{Jahanbakhshi, R.}} \yr{2021}  \at{Mechanisms of entrainment in a turbulent boundary layer}.  \jt{Phys. Fluids}  \bvol{33 \textnormal{(3)}},  \pg{035105}.

\bibitem[Jahanbakhshi \& Madnia(2016)]{Jahanbakhshi&Madnia2016}
{\sc \au{Jahanbakhshi, R.} \& \au{Madnia, C.~K.}} \yr{2016}  \at{Entrainment in a compressible turbulent shear layer}.  \jt{J. Fluid Mech.}  \bvol{797},  \pg{564--603}.

\bibitem[Jahanbakhshi \& Madnia(2018)]{Jahanbakhshi&Madnia2018}
{\sc \au{Jahanbakhshi, R.} \& \au{Madnia, C.~K.}} \yr{2018}  \at{Viscous superlayer in a reacting compressible turbulent mixing layer}.  \jt{J. Fluid Mech.}  \bvol{848},  \pg{743--755}.

\bibitem[Kankanwadi \& Buxton(2020)]{Kankanwadi&Buxton2020}
{\sc \au{Kankanwadi, K.~S.} \& \au{Buxton, O. R.~H.}} \yr{2020}  \at{Turbulent entrainment into a cylinder wake from a turbulent background}.  \jt{J. Fluid Mech.}  \bvol{905},  \pg{A35}.

\bibitem[Kankanwadi \& Buxton(2022)]{Kankanwadi&Buxton2022}
{\sc \au{Kankanwadi, K.~S.} \& \au{Buxton, O. R.~H.}} \yr{2022}  \at{On the physical nature of the turbulent/turbulent interface}.  \jt{J. Fluid Mech.}  \bvol{942},  \pg{A31}.

\bibitem[Kankanwadi \& Buxton(2023)]{Kankanwadi&Buxton2023}
{\sc \au{Kankanwadi, K.~S.} \& \au{Buxton, O. R.~H.}} \yr{2023}  \at{Influence of freestream turbulence on the near-field growth of a turbulent cylinder wake: Turbulent entrainment and wake meandering}.  \jt{Phys. Rev. Fluids}  \bvol{8 \textnormal{(3)}},  \pg{034603}.

\bibitem[Khorsandi {\em et~al.\/}(2013)Khorsandi, Gaskin \& Mydlarski]{Khorsandi2013}
{\sc \au{Khorsandi, B.}, \au{Gaskin, S.} \& \au{Mydlarski, L.}} \yr{2013}  \at{Effect of background turbulence on an axisymmetric turbulent jet}.  \jt{J. Fluid Mech.}  \bvol{736},  \pg{250--286}.

\bibitem[Kohan \& Gaskin(2020)]{Kohan&Gaskin2020}
{\sc \au{Kohan, K.~F.} \& \au{Gaskin, S.}} \yr{2020}  \at{The effect of the geometric features of the turbulent/non-turbulent interface on the entrainment of a passive scalar into a jet}.  \jt{Phys. Fluids}  \bvol{32 \textnormal{(9)}},  \pg{095114}.

\bibitem[Kohan \& Gaskin(2022)]{Kohan&Gaskin2022}
{\sc \au{Kohan, K.~F.} \& \au{Gaskin, S.~J.}} \yr{2022}  \at{On the scalar turbulent/turbulent interface of axisymmetric jets}.  \jt{J. Fluid Mech.}  \bvol{950},  \pg{A32 (referred to herein as KG22)}.

\bibitem[Kohan \& Gaskin(2023)]{Kohan&Gaskin2023}
{\sc \au{Kohan, K.~F.} \& \au{Gaskin, S.~J.}} \yr{2023}  \at{Scalar interfaces in the near field of a unity velocity ratio coaxial jet}.  \jt{Phys. Fluids}  \bvol{35 \textnormal{(3)}},  \pg{031711}.

\bibitem[Lai {\em et~al.\/}(2019)Lai, Law \& Adams]{Lai2019}
{\sc \au{Lai, A. C.~H.}, \au{Law, A. W-K.} \& \au{Adams, E.~E.}} \yr{2019}  \at{A second-order integral model for buoyant jets with background homogeneous and isotropic turbulence}.  \jt{J. Fluid Mech.}  \bvol{871},  \pg{271--304}.

\bibitem[Long {\em et~al.\/}(2022)Long, Wang \& Pan]{Long2022}
{\sc \au{Long, Y.}, \au{Wang, J.} \& \au{Pan, C.}} \yr{2022}  \at{Universal modulations of large-scale motions on entrainment of turbulent boundary layers}.  \jt{J. Fluid Mech.}  \bvol{941},  \pg{A68}.

\bibitem[Mi {\em et~al.\/}(2001)Mi, Nobes \& Nathan]{Mi2001}
{\sc \au{Mi, J.}, \au{Nobes, D.~S.} \& \au{Nathan, G.~J.}} \yr{2001}  \at{Influence of jet exit conditions on the passive scalar field of an axisymmetric free jet}.  \jt{J. Fluid Mech.}  \bvol{432},  \pg{91--125}.

\bibitem[Nakamura {\em et~al.\/}(2023)Nakamura, Watanabe \& Nagata]{Nakamura2023}
{\sc \au{Nakamura, K.}, \au{Watanabe, T.} \& \au{Nagata, K.}} \yr{2023}  \at{Turbulent/turbulent interfacial layers of a shearless turbulence mixing layer in temporally evolving grid turbulence}.  \jt{Phys. Fluids}  \bvol{35 \textnormal{(4)}},  \pg{045117}.

\bibitem[Parker {\em et~al.\/}(2019)Parker, Burridge, Partridge \& Linden]{Parker2019}
{\sc \au{Parker, D.~A.}, \au{Burridge, H.~C.}, \au{Partridge, J.~L.} \& \au{Linden, P.~F.}} \yr{2019}  \at{A comparison of entrainment in turbulent line plumes adjacent to and distant from a vertical wall}.  \jt{J. Fluid Mech.}  \bvol{882},  \pg{A4}.

\bibitem[Perez-Alvarado(2016)]{Perez-Alvarado2016}
{\sc \au{Perez-Alvarado, A.}} \yr{2016}  \at{Effect of background turbulence on the scalar field of a turbulent jet}. PhD thesis, McGill University.

\bibitem[Philip \& Marusic(2012)]{Philip&Marusic2012}
{\sc \au{Philip, J.} \& \au{Marusic, I.}} \yr{2012}  \at{Large-scale eddies and their role in entrainment in turbulent jets and wakes}.  \jt{Phys. Fluids}  \bvol{24 \textnormal{(5)}},  \pg{055108}.

\bibitem[Pope(2000)]{Pope2000}
{\sc \au{Pope, S.~B.}} \yr{2000} {\em Turbulent flows\/}.  \publ{Cambridge University Press}.

\bibitem[de~Rooy {\em et~al.\/}(2013)de~Rooy, Bechtold, Fr\"{o}hlich, Hohenegger, Jonker, Mironov, {Pier Siebesma}, Teixeira \& Yano]{deRooy2013}
{\sc \au{de~Rooy, W.~C.}, \au{Bechtold, P.}, \au{Fr\"{o}hlich, K.}, \au{Hohenegger, C.}, \au{Jonker, H.}, \au{Mironov, D.}, \au{{Pier Siebesma}, A.}, \au{Teixeira, J.} \& \au{Yano, J-I.}} \yr{2013}  \at{Entrainment and detrainment in cumulus convection: an overview}.  \jt{Q. J. R. Meteorol .Soc.}  \bvol{139 \textnormal{(670)}},  \pg{1--19}.

\bibitem[Sahebjam {\em et~al.\/}(2022)Sahebjam, Kohan \& Gaskin]{Sahebjam2022}
{\sc \au{Sahebjam, R.}, \au{Kohan, K.~F.} \& \au{Gaskin, S.}} \yr{2022}  \at{The dynamics of an axisymmetric turbulent jet in ambient turbulence interpreted from the passive scalar field statistics}.  \jt{Phys. Fluids}  \bvol{34},  \pg{015129}.

\bibitem[Shan \& Dimotakis(2006)]{Shan&Dimotakis2006}
{\sc \au{Shan, J.~W.} \& \au{Dimotakis, P.~E.}} \yr{2006}  \at{Reynolds-number effects and anisotropy in transverse-jet mixing}.  \jt{J. Fluid Mech.}  \bvol{566},  \pg{47--96}.

\bibitem[da~Silva {\em et~al.\/}(2014{\natexlab{{\em a\/}}})da~Silva, Hunt, Eames \& Westerweel]{AnnualReview2014}
{\sc \au{da~Silva, C.~B.}, \au{Hunt, J. C.~R.}, \au{Eames, I.} \& \au{Westerweel, J.}} \yr{2014{\natexlab{{\em a\/}}}}  \at{Interfacial layers between regions of different turbulence intensity}.  \jt{Ann. Rev. Fluid Mech.}  \bvol{46},  \pg{567--590}.

\bibitem[da~Silva {\em et~al.\/}(2014{\natexlab{{\em b\/}}})da~Silva, Taveira \& Borrell]{daSilva2014}
{\sc \au{da~Silva, C.~B.}, \au{Taveira, R.~R.} \& \au{Borrell, G.}} \yr{2014{\natexlab{{\em b\/}}}}  \at{Characteristics of the turbulent/nonturbulent interface in boundary layers, jets and shear-free turbulence}.  \jt{J. Phys.: Conf. Ser.}  \bvol{506 \textnormal{(1)}},  \pg{012015}.

\bibitem[Taveira {\em et~al.\/}(2013)Taveira, Diogo, Lopes \& da~Silva]{Taveira2013}
{\sc \au{Taveira, R.~R.}, \au{Diogo, J.~S.}, \au{Lopes, D.~S.} \& \au{da~Silva, C.~B.}} \yr{2013}  \at{Lagrangian statistics across the turbulent-nonturbulent interface in a turbulent plane jet}.  \jt{Phys. Rev. E.}  \bvol{88 \textnormal{(4)}},  \pg{043001}.

\bibitem[Westerweel {\em et~al.\/}(2009)Westerweel, Fukushima, Pedersen \& Hunt]{Westerweel2009}
{\sc \au{Westerweel, J.}, \au{Fukushima, C.}, \au{Pedersen, J.~M.} \& \au{Hunt, J. C.~R.}} \yr{2009}  \at{Momentum and scalar transport at the turbulent/non-turbulent interface of a jet}.  \jt{J. Fluid Mech.}  \bvol{631},  \pg{199--230}.

\bibitem[Wu {\em et~al.\/}(2020)Wu, Wang \& Pan]{Wu2020}
{\sc \au{Wu, D.}, \au{Wang, J.} \& \au{Pan, C.}} \yr{2020}  \at{Bubbles and drops in the vicinity of turbulent/non‑turbulent interface in turbulent boundary layers}.  \jt{Exp. Fluids}  \bvol{61 \textnormal{(11)}},  \pg{1--11}.

\bibitem[Wu {\em et~al.\/}(2019)Wu, Wallace \& Hickey]{Wu2019}
{\sc \au{Wu, X.}, \au{Wallace, J.~M.} \& \au{Hickey, J.~P.}} \yr{2019}  \at{Boundary layer turbulence and freestream turbulence interface, turbulent spot and freestream turbulence interface, laminar boundary layer and freestream turbulence interface}.  \jt{Phys. Fluids}  \bvol{31 \textnormal{(4)}},  \pg{045104}.

\bibitem[Xu {\em et~al.\/}(2023)Xu, Long \& Wang]{Xu2023}
{\sc \au{Xu, C.}, \au{Long, Y.} \& \au{Wang, J.}} \yr{2023}  \at{Entrainment mechanism of turbulent synthetic jet flow}.  \jt{J. Fluid Mech.}  \bvol{958},  \pg{A31}.

\bibitem[You \& Zaki(2019)]{You&Zaki2019}
{\sc \au{You, J.} \& \au{Zaki, T.~A.}} \yr{2019}  \at{Conditional statistics and flow structures in turbulent boundary layers buffeted by free-stream disturbances}.  \jt{J. Fluid Mech.}  \bvol{866},  \pg{526--566}.

\bibitem[Yule(1978)]{Yule1978}
{\sc \au{Yule, A.~J.}} \yr{1978}  \at{Large-scale structure in the mixing layer of a round jet}.  \jt{J. Fluid Mech.}  \bvol{89 \textnormal{(3)}},  \pg{413--432}.

\end{thebibliography}
\bibliographystyle{jfm}

\end{document}